\documentclass[a4paper,11pt]{article}
\usepackage{graphics,graphicx,epsfig,wrapfig}
\usepackage{longtable}
\usepackage{amsmath,verbatim,amssymb,color,lscape}
\usepackage{kotex}
\usepackage{natbib}
\usepackage{lastpage}
\usepackage{multirow}
\usepackage{authblk}
\usepackage{bm}
\usepackage{indentfirst}
\usepackage[ruled,vlined]{algorithm2e}
\usepackage{algorithmic}
\usepackage[normalem]{ulem}
\usepackage[export]{adjustbox}
\usepackage{verbatim}
\usepackage{tabularx}
\usepackage{booktabs}
\usepackage{siunitx}
\usepackage{url}

\newcommand{\hide}[1]{}

\makeatletter

\usepackage[outerbars,color]{changebar}
\ifx\pdfoutput\undefined
\else\ifnum\pdfoutput>0
  \usepackage{pdfcolmk}
\fi\fi
\cbcolor{black}

\usepackage[margin=1.0in]{geometry}
\usepackage{tikz}
\usetikzlibrary{bayesnet}
\usetikzlibrary{fit,positioning}

\newfont{\rmm}{cmr10 at 11pt}
\rmm

\title{A Representation-Learning Item Response Model for Identifying Behaviorally Important Actions in PIAAC Process Data}

\author[1]{Junyeong Park}
\author[1]{Daeun Hwangbo}
\author[1]{Seyoung Park}
\author[1,2]{Ick Hoon Jin}
\author[3]{Minjeong Jeon}
\affil[1]{Department of Statistics and Data Science, Yonsei University. Republic of Korea.}
\affil[2]{Department of Applied Statistics, Yonsei University. Republic of Korea.}
\affil[3]{School of Education and Information Studies, University of California, Los Angeles. USA.}
\date{}

\begin{document}
\maketitle

\begin{abstract}
Problem-solving log process data from computer-based assessments provide detailed information about how respondents approach and complete tasks. However, the resulting action sequences are complex and noisy, making it difficult to identify specific behaviors associated with successful performance. This paper proposes a representation-learning item response modeling (IRT) framework for identifying behaviorally important actions while accounting for respondent proficiency and item-level differences. Raw log sequences and timing information are first transformed into action representations that incorporate the hierarchical structure of action labels and the sequential and temporal context in which each action occurs. These respondent-specific representations are then entered as covariates in an extended IRT model, with spike-and-slab priors used to identify action–item combinations associated with response accuracy. The framework therefore evaluates actions contextually rather than as simple occurrence indicators and provides posterior uncertainty for their associations with performance. We apply the approach to problem-solving process data from the OECD Programme for the International Assessment of Adult Competencies (PIAAC). The analysis identifies a sparse set of actions associated with successful and unsuccessful performance and reveals differences across items in where behavioral information occurs within the problem-solving process. 
\end{abstract}

\noindent {\bf Keywords:} Process Data, Item Response Theory, Behavioral Embedding, Spike-and-slab Prior, Problem-solving Assessment
\newpage

\section{Introduction}\label{sec:intro}

Large-scale computer-based assessments increasingly record how respondents interact with the test interface. In the problem solving in technology-rich environments (PSTRE) domain of the OECD Programme for the International Assessment of Adult Competencies (PIAAC), respondents complete tasks in simulated digital environments, and each discrete action, such as page navigation, drag-and-drop operations, and form submissions, is stored as a timestamped event \citep{oecd2012literacy, piaac2009pstre, organisation2019beyond, goldhammer2020analysing}. In the Party Invitations item of PIAAC, for example, respondents read email replies to an invitation, file each message into a folder labeled \textit{Can Come} or \textit{Cannot Come}, and submit the resulting classification. The log records the ordered sequence of views, drags, and drops that produced the final answer. Process data of this kind capture the problem-solving course rather than only its outcome \citep{kroehne2018conceptualize, goldhammer2021byproduct, maddox2023uses, stadler2020first} and respondents who answer correctly may traverse systematically different interaction sequences than those who answer incorrectly \citep{eichmann2020exploring, he2021leveraging}. Identifying the behaviors that distinguish successful from unsuccessful problem solving is therefore a central goal of process data analysis. Such behaviors indicate where instructional feedback should focus, inform task design, and improve the interpretation of assessment results \citep{zoanetti2010interactive, zhang2023identifying, zhang2023accurate}.

Much existing work relates behavior to response outcomes by comparing correct and incorrect respondents. Early studies compared analyst-defined summary features such as action counts and time on task \citep{goldhammer2014time, greiff2016understanding}. Attention then turned to behavioral patterns mined from the raw sequences \citep{he2015identifying, he2016analyzing, ulitzsch2022using}, and a more recent line selects individual actions \citep{zhou2024investigating, yun2025discovering}. These approaches generally identify actions from their associations with response outcomes outside a measurement model that explicitly accounts for respondent proficiency and item-level differences. An action may appear discriminating partly because respondents of higher proficiency tend to perform it. More generally, associations between actions and response accuracy can be confounded with respondent- and item-level differences unless these sources of heterogeneity are represented explicitly. Our interest is therefore in whether an action provides information about response success after accounting for systematic differences among respondents and items. Item response models provide a natural measurement framework for representing these sources of heterogeneity. Measurement models for process data supply this conditioning but have targeted the global structure of the process, such as response speed or transition dynamics, rather than individual actions \citep{van2007hierarchical, wang2015mixture, chen2020continuous, xu2020latent, park2025analysis}. 

We therefore bring action-level selection into an item response model, allowing the association between each contextualized action representation and response accuracy to be evaluated while accounting for respondent proficiency and item-level differences. This approach raises three statistical challenges. First, the covariates are not given. Raw logs are variable-length sequences of discrete events with noisy, hierarchically structured labels, and the meaning of an event depends on its context and timing. The same action may be exploratory early in an episode and confirmatory near its end. Thus, treating an action simply as present or absent may overlook substantial variation in the behavioral context in which that action occurs. Representations of actions must therefore capture not only action identity but also the context and timing of its occurrence. Manual feature engineering has filled this role but is labor intensive, task specific, and prone to discarding sequential information, which has motivated data-driven representation learning for process data \citep{tang2020latent, tang2021exploratory, zhou2024investigating, scarlatos2022process}. Second, the predictor space is large relative to the sample. Our analysis involves 2,025 unique action--item combinations among 1,996 respondents, and only a small fraction of the actions is expected to matter. Accordingly, selection must be sparse and must quantify uncertainty. Third, the conditioning quantities are themselves unknown. Ability and difficulty are latent and must be estimated from the same responses, which places selection inside a logistic model with crossed respondent and item effects.

We address these challenges with a three-stage framework that couples representation learning with Bayesian variable selection. In the first stage, embedding methods from natural language processing \citep{mikolov2013efficient, mikolov2013distributed, bojanowski2017enriching} map each action to a dense vector that reflects co-occurrence in similar sequential contexts and shared components in the hierarchical event labels. In the second stage, a long short-term memory (LSTM) autoencoder \citep{hochreiter1997long, srivastava2015unsupervised} compresses the embedded and temporally augmented sequence of each respondent into a low-dimensional latent value for every action occurrence. The resulting covariate is respondent specific, encoding both the action and the sequential context and timing of its performance. In the third stage, the latent values enter an extended Rasch model as covariates whose weights receive spike-and-slab priors \citep{mitchellbeauchamp1988, george1993variable, ishwaran2005spike, chen2022advantages}. Estimation and selection proceed simultaneously through Markov chain Monte Carlo, and each action--item combination receives a posterior inclusion probability and a posterior effect distribution defined conditional on ability and difficulty. Thus, the learned covariates represent not simply whether an action occurred, but how that action occurred within the respondent's sequential and temporal problem-solving context. The selection model then evaluates whether this contextualized behavioral information is associated with response success after accounting for respondent and item effect.

This paper makes three contributions. First, we introduce an action-level extension of an item response model that evaluates associations between problem-solving behaviors and response accuracy while accounting for respondent proficiency and item-level differences. This places the identification of behaviorally important actions within a common measurement framework rather than relying only on marginal contrasts between performance groups. Second, we represent actions contextually rather than as simple occurrence indicators. Representation learning combines the hierarchical structure of action labels with their sequential and temporal context, allowing the representation of the same action to vary according to how and when it occurs in a respondent's solution process. Third, we incorporate sparse Bayesian selection into this framework, providing posterior inclusion probabilities and effect distributions for action–item combinations rather than only ranked importance scores. Applied to PIAAC PSTRE data, the approach identifies sparse sets of behaviors associated with successful and unsuccessful problem solving and reveals substantial differences across items in where behavioral information occurs within the solution process.

The remainder of the paper is organized as follows. Section~\ref{sec:related_work} reviews the PIAAC PSTRE data and related work. Section~\ref{sec:Pre_process} describes the construction of analysis-ready action sequences from the raw logs. Section~\ref{sec:methodology} presents the proposed methodology. Section~\ref{sec:real_data} reports the empirical results. Section~\ref{sec:simul} evaluates the variable-selection performance through a simulation study. Section~\ref{sec:conc} concludes with implications and limitations.

\section{Background and Related Work}\label{sec:related_work}
\subsection{PIAAC PSTRE and Process Data}\label{sec:piaac_data}

PIAAC is an international assessment developed by the OECD to measure information-processing skills among adults aged 16--65. Its first cycle was conducted in 2011--2012 \citep{oecd2013piaac}. The assessment covers three domains, literacy, numeracy, and PSTRE. In the PSTRE domain, respondents work in simulated email clients, web browsers, and spreadsheets to complete problem-solving tasks \citep{piaac2009pstre} and the platform records each interaction with the interface. The log files trace how a respondent approached and executed a task rather than only whether the response was correct.

\begin{table}[!h]
\centering
\begin{tabular}{llcr}
\toprule
Item & Task & Response type & $N_j$ \\
\midrule
PS1-1 & Party Invitations (Part~1) & Polytomous (0--3) & 1,295 \\
PS1-2 & Party Invitations (Part~2) & Binary            & 1,246 \\
PS1-3 & CD Tally                   & Binary            & 1,272 \\
PS1-4 & Sprained Ankle (Part~1)    & Binary            & 1,255 \\
PS1-5 & Sprained Ankle (Part~2)    & Binary            & 1,302 \\
PS1-6 & Tickets                    & Binary            & 1,282 \\
PS1-7 & Class Attendance           & Polytomous (0--3) & 1,074 \\
\midrule
PS2-1 & Club Membership (Part~1)   & Binary            & 1,274 \\
PS2-2 & Club Membership (Part~2)   & Polytomous (0--3) & 1,170 \\
PS2-3 & Book Order                 & Binary            & 1,238 \\
PS2-4 & Meeting Room               & Polytomous (0--3) & 1,161 \\
PS2-5 & Reply All                  & Binary            & 1,199 \\
PS2-6 & Locate Email               & Polytomous (0--3) & 1,131 \\
PS2-7 & Lamp Return                & Polytomous (0--3) & 1,230 \\
\bottomrule
\end{tabular}
\caption{PSTRE items used in the analysis, administered in two blocks, PS1 and PS2. $N_j$ denotes the number of respondent--item pairs for item $j$ with non-empty action sequences after preprocessing.}
\label{tab:items}
\end{table}

The raw log data are organized at the event level. Each row records a single event and contains the respondent sequence identifier (\path{SEQID}), the item identifier, the event name, the event type, a timestamp, and an event description. The field \path{event_type} gives a broad action category such as \path{MAIL_DROP}, \path{FOLDER_VIEWED}, or \path{TOOLBAR}, and \path{event_description} supplies contextual detail about the object or interface element involved, for instance which message was moved and into which folder. A single user action may generate several log entries, combining the action itself with ancillary events generated by the system. Constructing analysis-ready action sequences therefore requires preprocessing \citep{hwangbo2026tutorial}, described in Section~\ref{sec:preprocessing_pipeline}.

The present analysis uses the United States sample from the first cycle and the 14 PSTRE items administered in two blocks, PS1 and PS2. The items span a range of task complexity. Simpler items require a single application, as when respondents read a webpage and select a response option. More complex items require navigation across applications and integration of information from several sources. The Lamp Return item (PS2-7), for example, requires respondents to navigate a multi-page website, complete an online exchange form, and confirm the transaction. Items are scored either as binary correct or incorrect outcomes or on a polytomous scale from 0 to 3. Log files and scored outcomes are available for 1,996 respondents. Respondents did not all receive the same items, and the number of respondent--item pairs retained after preprocessing ranges from 1,074 to 1,302 across items. Table~\ref{tab:items} lists the items, their scoring formats, and these counts.

\subsection{Prior Approaches to Process Data Analysis}\label{sec:pre_process_data}

Early work on process data constructed scalar summary statistics from the log sequences, including the number of actions, total time on task, and indicators for specific event types, and entered these summaries into regression or latent variable models \citep{goldhammer2014time, goldhammer2020analysing, lundgren2020within, costa2023exploring, greiff2016understanding}. Summaries of this kind are straightforward to compute and to interpret. Aggregation over the sequence, however, records the amount of activity while discarding which actions were performed and in what order.

A second strand extracts behavioral structure from the raw sequences without analyst-defined indicators. Methods include $n$-gram pattern analysis \citep{he2015identifying, he2016analyzing}, multidimensional scaling of variable-length sequences \citep{tang2020latent}, sequence-to-sequence autoencoder representations \citep{tang2021exploratory}, longest-common-subsequence identification of behavioral paths \citep{he2019using, he2021leveraging}, clustering and network analysis of response processes \citep{ulitzsch2021combining, zhang2023identifying}, path signature features that encode joint action and timing structure \citep{tang2025path}, and latent class and neural network representations \citep{fang2020latent, wang2023subtask, zhou2024investigating}. These studies establish that raw sequences carry behavioral information beyond what scalar summaries capture. Applications have contrasted the sequences of correct and incorrect respondents at the level of patterns \citep{stadler2019taking, ulitzsch2022using} and have used the learned representations for prediction \citep{wang2023subtask} and for the description of behavioral patterns \citep{tang2020latent, zhang2023identifying, zhou2024investigating}. Findings are stated at the level of patterns or respondent groups rather than individual actions.

A separate line of work selects key actions directly \citep{zhou2024investigating, yun2025discovering}. Selection there is applied uniformly across items and rests on the actions themselves, without reference to respondent ability or item difficulty, and the selected actions may therefore reflect differences in respondent ability rather than information specific to the action. Item response models provide the conditioning that such selection lacks. They express the log-odds of a correct response through a respondent ability parameter and an item difficulty parameter estimated jointly from the observed responses, and the explanatory item response tradition shows that auxiliary predictors can be entered into this expression and estimated conditional on both \citep{deboeck2004explanatory}. Process information has been brought into this framework in three ways. The first treats response time as a second outcome and links it to accuracy through a hierarchical structure over respondents and items \citep{van2007hierarchical, wang2015mixture}. The second models the action process itself, through continuous-time choice, latent topic, and sequential response formulations \citep{chen2020continuous, xu2020latent, han2022sequential, park2025analysis}. The third extracts features from the log sequences and enters them into item response models to predict response accuracy and to sharpen the estimation of latent ability \citep{scarlatos2022process, zhang2023accurate}. In all three, the process enters at the level of the whole sequence, as transition dynamics, response speed, or sequence-level features, and no individual action is selected. The framework developed here places action-level selection inside a model of this type. Learned action representations, reviewed in Section~\ref{sec:pre_repre}, enter an extended Rasch model in which a spike-and-slab prior \citep{mitchellbeauchamp1988, ishwaran2005spike} identifies the actions associated with accuracy conditional on respondent ability and item difficulty.

\subsection{Representation Learning for Sequential Data}\label{sec:pre_repre}

Word2Vec learns dense vector representations of discrete symbols from their distributional context. The skip-gram model predicts the symbols surrounding a center symbol within a fixed window, and symbols appearing in similar contexts receive similar vectors \citep{mikolov2013efficient, mikolov2013distributed}. FastText represents each word as the sum of word-level and subword-level embeddings \citep{bojanowski2017enriching}, which allows related words to share statistical strength and yields representations for rare words. The latter property matters when the vocabulary is large and individual symbols are infrequent.

These methods have been adapted beyond natural language, to purchase histories and item interaction data \citep{grbovic2015ecommerce, barkan2016item2vec} and to enrollment sequences in educational data mining \citep{pardos2020university}. Closest to the present setting, Word2Vec embeddings have been applied to action sequences from large-scale problem-solving assessments to extract behavioral information from raw logs and to identify actions that distinguish performance groups \citep{zhou2024investigating, yun2025discovering}.

Long short-term memory (LSTM) networks model long-range dependencies in sequential data through gated memory cells \citep{hochreiter1997long}. An LSTM autoencoder compresses a variable-length sequence into a latent representation by reconstructing the input from an encoded bottleneck, without predefined feature extraction \citep{srivastava2015unsupervised, malhotra2016lstm}. Whereas $n$-gram features encode the local co-occurrence of adjacent actions, the recurrent structure allows information from the entire sequence to inform the representation at each position. This property motivates the use of an LSTM autoencoder in the second stage of the proposed framework.

\section{Action Unit Construction and Preprocessing}\label{sec:Pre_process}
\subsection{Unit and Token Definitions}\label{sec:unit_token}

The modeling unit in this study is the \emph{action unit}, a standardized label representing one event in the problem-solving sequence of a respondent. Each action unit combines the event category recorded in \path{event_type} with a cleaned portion of the contextual information recorded in \path{event_description}, when available. The label retains the type of interface operation performed together with the task-specific detail that distinguishes instances of the same operation.

A unit label consists of semantic components connected by underscores, referred to as \emph{action tokens}. In \path{mail-drop_cancome-folder}, the tokens \path{mail-drop} and \path{cancome-folder} correspond to the operation type and the target object.

The distinction between units and tokens matters for the embedding model of Section~\ref{sec:hw2v} for two reasons. First, it mirrors the generative structure of the log labels. Event labels arise from a hierarchical scheme in which an operation category (\path{event_type}) is qualified by lower-level descriptors of the object or location (\path{event_description}), and treating each fully qualified label as atomic would discard this hierarchy. Second, shared tokens carry semantic relatedness. The units \path{mail-drop_cancome-folder} and \path{mail-drop_cannotcome-folder} denote the same operation on different targets, whereas \path{mail-drop_cancome-folder} and \path{folder-viewed_cancome-folder} denote different operations on the same target. Units define the vocabulary for learning sequential co-occurrence, and tokens supply components through which related units share statistical strength, which gives rare units an informed representation from few direct occurrences. This structure motivates the FastText-style subword composition adopted in Section~\ref{sec:hw2v}.

\subsection{Preprocessing Pipeline}\label{sec:preprocessing_pipeline}

The raw log files contain events that are too granular, redundant, or system-dependent to serve as modeling units. We preprocess the logs into consistent action sequences while preserving temporal order and behavioral distinctions. The procedure builds on the pipeline of \citet{hwangbo2026tutorial}, which combines rule-based cleaning of timestamp reversals, duplicate records, and multi-entry actions with an LLM-assisted implementation workflow. We organize the pipeline into three stages, rule-based cleaning, LLM-assisted description standardization, and action unit construction. The preprocessing scripts are available at \url{https://github.com/P-JuNYeonG/action-irt}.

The first stage consists of rule-based preprocessing. This stage addresses three categories of issues: timestamp reversals, duplicate records, and multiple log entries per user action. Restart events require special handling because the system-generated restart resets the timestamp to zero in the middle of a sequence. Following the offset-correction procedure described in \citet{hwangbo2026tutorial}, timestamps recorded after a restart are adjusted by adding a restart-specific offset so that the full sequence remains temporally ordered on a single cumulative time scale. Duplicate records, defined as log entries that share the same event type and timestamp within the same respondent--item pair, are removed. Events that are administrative or system-generated rather than respondent-initiated are also excluded, including the session-start marker recorded at item presentation and the session-end marker recorded at item completion. When a single user action generates multiple log entries comprising a core action and ancillary events, these entries are consolidated into a single representative action that retains the analytically relevant information. Finally, consecutive \path{KEYPRESS} events are collapsed into a single keypress action whose label records the number of consecutive keypresses, preventing long runs of individual keystrokes from dominating the action sequence while preserving information about the extent of text entry. Action labels that recur across items with inconsistent or uninformative raw identifiers are also standardized at this stage. Submission-related interface elements described differently across items were mapped to consistent labels such as ``next'', ``OK'', and ``cancel''.

The second stage addresses the event descriptions that remain after rule-based cleaning. The \path{event_description} field retains system-specific identifiers, numerical indices, and delimiter-separated strings that are too noisy to use as components of action unit labels. These descriptions are normalized into substitutes that preserve the distinguishing information. A raw description containing internal element identifiers, for example, is reduced to a label retaining the target object or interface location. The normalization requires context-dependent interpretation of strings across a heterogeneous set of descriptions, and we employ an LLM-assisted procedure adapted from the prompt-based workflow of \citet{hwangbo2026tutorial}. The procedure is applied within event-type groups, and descriptions from different action categories are processed independently. Section~\ref{sec:llm_preprocessing} describes the prompt design, the choice of model, and the adaptations made for the present study.

The third stage combines the cleaned event type and the cleaned description into the final action unit following the construction rule of Section~\ref{sec:unit_token}. If the cleaned description contains information, it is concatenated with the event type using underscores; otherwise the event type alone is retained as the unit label. The resulting data for each item consist of respondent-level ordered sequences of action units and their timestamps, which serve as input to the action embedding model of Section~\ref{sec:hw2v}.

\subsection{LLM-Assisted Description Standardization}\label{sec:llm_preprocessing}

The LLM-assisted step normalizes the event descriptions that remain after rule-based cleaning. Its scope is restricted to string-level normalization and excludes inference about respondent strategies and substantive analytic decisions.

The input to the LLM is organized as groups of descriptions within each \path{event_type}. For each group, the model receives the set of raw description strings and returns a table containing the original event type, the original description, and a proposed substitute that retains the distinguishing information. Several constraints keep the transformation conservative. Task-specific identifiers are preserved as single semantic units, redundant components are removed only when they duplicate information present elsewhere in the label, and descriptions that do not form a coherent group are retained in their original form. The transformation rules, constraint specifications, and prompt text are provided in Section~B of the Supplementary Material.

After the substitutes are produced, the researcher reviews each mapping and connects the cleaned substitute values to the corresponding event types to form final action units. This human-in-the-loop design uses the LLM for normalization of a heterogeneous set of description strings while keeping the final preprocessing decisions auditable and under researcher control. In the present analysis, Claude Sonnet 4.5 \citep{anthropic2025sonnet} was used for this step, accessed via the OpenRouter API. The final researcher-verified mapping table is included in the project's GitHub repository, allowing the preprocessing outcome to be reproduced directly from the stored mapping without re-executing the LLM step.

\section{Methodology}\label{sec:methodology}

\begin{figure}[htb!]
    \centering
    \includegraphics[width=0.9\linewidth]{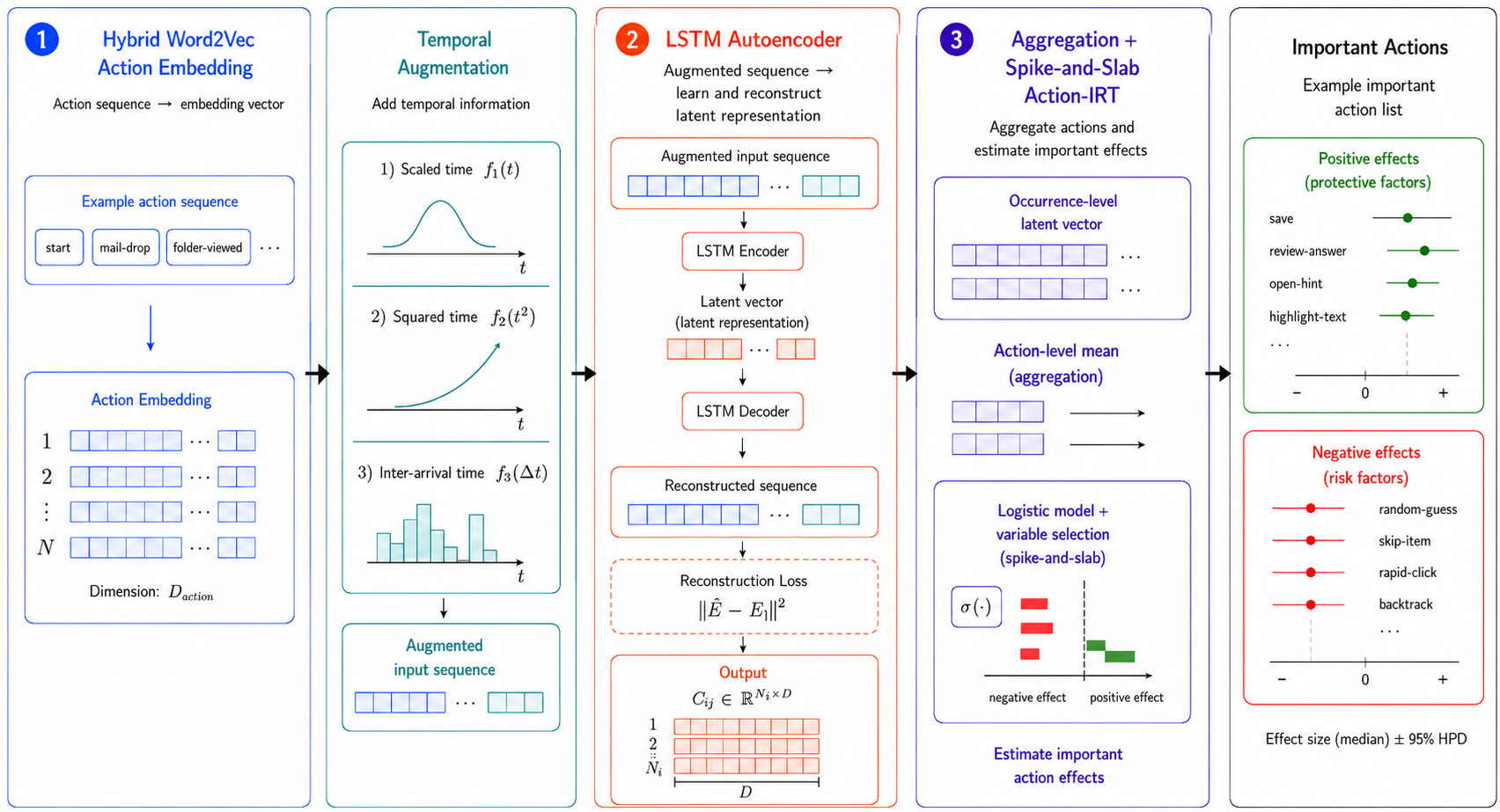}
    \caption{
    Schematic overview of the three-stage framework. In Stage~1, the action-unit sequence of respondent~$i$ on item~$j$ is embedded by a hybrid Word2Vec model and augmented with three temporal features, yielding $\tilde{\mathbf{E}}_{ij} \in \mathbb{R}^{N_{ij} \times (D_{\mathrm{Action}}+3)}$ for the $N_{ij}$ actions in the sequence. In Stage~2, an LSTM autoencoder compresses the augmented embeddings into a $D$-dimensional latent value for each action occurrence, averaged over repeated occurrences to give $\bar{C}_{ijl}^{(d)}$ for unique action~$l$. In Stage~3, these values enter the action-IRT model as fixed covariates whose weights receive a spike-and-slab prior. Actions are selected through posterior inclusion probabilities, and the posterior action effect $\Delta_{jl}$ gives the direction and magnitude of the contribution of action~$l$ to the log-odds of a correct response on item~$j$. 
    }
    \label{fig:framework}
\end{figure}

We propose a three-stage framework for identifying important actions in problem-solving process data. An action is important for item~$j$ if its latent representation is associated with response accuracy conditional on respondent ability and item difficulty. The first stage maps each action unit to a dense continuous vector through a hybrid Word2Vec model \citep{mikolov2013efficient, mikolov2013distributed} that uses the sequential context of action units together with the compositional structure of their constituent tokens. The second stage compresses the temporally augmented embedding sequence of each respondent into a low-dimensional latent representation through an LSTM autoencoder \citep{hochreiter1997long, srivastava2015unsupervised}. The third stage enters the latent action representations into an extended Rasch model \citep{rasch1960probabilistic} with a spike-and-slab prior \citep{mitchellbeauchamp1988, george1993variable, ishwaran2005spike}, which performs variable selection over the action space through MCMC-based posterior inference. Figure~\ref{fig:framework} provides a schematic overview. Throughout, $D_{\mathrm{Action}}$ denotes the dimension of the action-unit embedding produced in Stage~1, and $D$ denotes the dimension of the latent representation produced by the LSTM autoencoder in Stage~2. The two dimensions are set independently, with $D \leq D_{\mathrm{Action}}$ in practice.

Let $i=1,\ldots,N$ index respondents and $j=1,\ldots,J$ index items. For respondent~$i$ and item~$j$, let $Y_{ij}\in\{0,1\}$ denote the binary response outcome, where $Y_{ij}=1$ indicates a correct response. The preprocessed action sequence for the pair $(i,j)$ is
\[
\mathcal{A}_{ij}=(a_{ij1},a_{ij2},\ldots,a_{ijN_{ij}}),
\]
where $N_{ij}$ is the length of the sequence and each $a_{ijn}$ is an action unit. As described in Section~\ref{sec:unit_token}, each action unit decomposes into lower-level action tokens that reflect the hierarchical structure of the log event label. For item~$j$, let $L_j$ denote the number of unique action units observed across all respondents, and let $A_{ij}\subseteq\{1,\ldots,L_j\}$ be the set of unique action indices appearing in $\mathcal{A}_{ij}$. For each unique action $l \in A_{ij}$, let $R_{ijl}$ denote the number of times action~$l$ appears in $\mathcal{A}_{ij}$, with $\sum_{l \in A_{ij}} R_{ijl} = N_{ij}$.

The framework produces a latent action representation $C_{ijn}^{(d)}\in\mathbb{R}$ for each action occurrence~$n$ and latent dimension $d=1,\ldots,D$. When a unique action~$l$ occurs $R_{ijl}$ times in the sequence of respondent~$i$ for item~$j$, we summarize its representation by the within-sequence average
\[
\bar{C}_{ijl}^{(d)}
=\frac{1}{R_{ijl}}\sum_{k=1}^{R_{ijl}} C_{ijlk}^{(d)},
\quad d=1,\ldots,D,
\]
where $C_{ijlk}^{(d)}$ is the $d$th latent value of the $k$th occurrence of action~$l$. The pair $(i,j)$ identifies a respondent--item pair, the subscript~$l$ identifies a unique action within item~$j$, and the superscript~$(d)$ identifies a coordinate of the $D$-dimensional latent representation. The quantity $\bar{C}_{ijl}^{(d)}$ is a respondent-specific summary of how action~$l$ was performed.

\subsection{Hybrid Word2Vec for Action Embedding}\label{sec:hw2v}

The first stage converts each action unit into a dense continuous vector. We assume that the action sequence $\mathcal{A}_{ij}$ exhibits approximate Markov dependence, in that the conditional distribution of each action given the sequence history depends primarily on the actions within a surrounding context window. Under this assumption, we adapt the skip-gram objective of Word2Vec \citep{mikolov2013efficient, mikolov2013distributed} to action sequences, training a center action unit to predict its neighboring units. Actions appearing in similar local contexts thereby receive similar vector representations.

Each preprocessed action unit has an internal hierarchical structure. As described in Section~\ref{sec:unit_token}, it decomposes into lower-level action tokens corresponding to semantic components of the event label. The unit \path{mail-drop_cancome-folder}, for example, decomposes into tokens encoding the mail drop operation and the target folder. Different units may share one or more tokens while differing in others, and a purely unit-level embedding would leave this partial overlap unused.

To preserve this compositional structure, we adopt a FastText-style subword composition \citep{bojanowski2017enriching}. The model maintains separate embedding vectors for action units and action tokens. For an action unit~$u$ composed of $m$ tokens $g_1, \ldots, g_m$, the composite representation is
\[
\mathbf{e}(u) = \mathbf{v}_{u}+\sum_{r=1}^{m}\mathbf{v}_{g_r},
\]
where $\mathbf{v}_{u}\in\mathbb{R}^{D_{\mathrm{Action}}}$ is the unit-level embedding and $\mathbf{v}_{g_r} \in \mathbb{R}^{D_{\mathrm{Action}}}$ is the embedding of token~$g_r$. The additive composition allows rare action units that share tokens with more frequent units to benefit from the token-level training signal, which mitigates data sparsity.

Training pairs are of two types. In \emph{unit--unit} pairs, a center unit is paired with each neighboring unit within the context window. In \emph{token--unit} pairs, each token of the center unit is paired with the same neighbors. Both types are trained jointly under a skip-gram objective with negative sampling, which draws units that occur in similar sequential contexts or share tokens toward similar representations. The training objective, negative-sampling procedure, and hyperparameter settings are provided in Section~A of the Supplementary Material.

After training, each action occurrence is represented by a vector in $\mathbb{R}^{D_{\mathrm{Action}}}$, and the sequence for respondent~$i$ on item~$j$ yields an embedding matrix $\mathbf{E}_{ij} \in \mathbb{R}^{N_{ij} \times D_{\mathrm{Action}}}$. This representation captures sequential and compositional structure without recording when each action occurred, and we augment each embedding vector with three temporal features. Let $t_{ijn}$ denote the elapsed time from the start of the episode to the $n$th action, and let $f_j(\cdot)$ denote standardization to zero mean and unit variance computed across all action occurrences of item~$j$. The augmented embedding vector for the $n$th action is
\[
\tilde{\mathbf{e}}_{ijn} = \bigl[ \mathbf{e}(a_{ijn}),\; f_j(t_{ijn}),\; f_j(t_{ijn}^{2}),\; f_j(t_{ijn}-t_{ij,n-1}) \bigr],
\]
where $t_{ij0} = 0$ by convention. The three temporal components capture the elapsed time of the action within the episode, nonlinear timing effects, and the local inter-action interval. Item-level standardization places the temporal features on a numerical scale comparable to the embedding coordinates. The resulting augmented sequence matrix is
\[
\tilde{\mathbf{E}}_{ij}\in\mathbb{R}^{N_{ij}\times(D_{\mathrm{Action}}+3)}.
\]

\subsection{LSTM Autoencoder for Dimension Reduction}\label{sec:lstm_ae}

The second stage reduces each augmented action embedding vector to a low-dimensional latent value that serves as an action-level covariate in the item response model. Two considerations motivate the reduction. First, the $(D_{\mathrm{Action}}+3)$-dimensional augmented embedding is too high-dimensional to enter the response model directly, since a separate weight for each coordinate of every unique action would introduce more parameters than the sample size supports. Second, the skip-gram embedding and its temporal augmentation encode the identity and timing of each action without encoding the sequential context in which it occurs. Individual actions may satisfy the local dependence assumed by the context window of Section~\ref{sec:hw2v}, while the problem-solving process extends over a longer sequence that a local window cannot represent, and the same action may carry different behavioral significance in an early exploratory phase and in a later confirmatory phase. A reduction that processes the full ordered sequence can encode this context into each per-action representation. An LSTM autoencoder \citep{hochreiter1997long, srivastava2015unsupervised} meets both requirements. The recurrent encoder summarizes the history of preceding actions at every position, and the reconstruction objective provides a training criterion that requires no response labels.

For each item~$j$, the input to the LSTM autoencoder is $\{\tilde{\mathbf{E}}_{ij}\}_{i=1}^{N}$. Let $H_{\mathrm{enc}}$ denote the hidden-state dimension of the encoder. The encoder processes the sequence $(\tilde{\mathbf{e}}_{ij1}, \ldots, \tilde{\mathbf{e}}_{ijN_{ij}})$ step by step and produces a hidden state $\mathbf{h}_{ijn} \in \mathbb{R}^{H_{\mathrm{enc}}}$ at each position~$n$. A linear projection maps the hidden state to the $D$-dimensional latent space,
\[
\mathbf{C}_{ijn} = \mathbf{W}_{\mathrm{proj}}\,\mathbf{h}_{ijn} + \mathbf{b}_{\mathrm{proj}} \in \mathbb{R}^{D},
\]
where $\mathbf{W}_{\mathrm{proj}} \in \mathbb{R}^{D \times H_{\mathrm{enc}}}$ and $\mathbf{b}_{\mathrm{proj}} \in \mathbb{R}^{D}$ are learnable parameters. The $d$th element of $\mathbf{C}_{ijn}$ is $C_{ijn}^{(d)}$ as defined in Section~\ref{sec:methodology}. Because $\mathbf{h}_{ijn}$ depends on all preceding inputs $\tilde{\mathbf{e}}_{ij1}, \ldots, \tilde{\mathbf{e}}_{ijn}$, the latent vector $\mathbf{C}_{ijn}$ encodes the $n$th action together with the sequential context in which it was performed. This property distinguishes the reduction from context-free alternatives such as principal component analysis or a standard autoencoder applied to each action vector independently \citep{hinton2006reducing}.

The decoder maps each latent vector back through a symmetric LSTM architecture to reconstruct the augmented embedding. Let $\hat{\mathbf{e}}_{ijn}$ denote the reconstructed vector at position~$n$. The training criterion is the mean squared reconstruction error over observed positions,
\[
\mathcal{L}_{j} = \frac{1}{\sum_{i} N_{ij}} \sum_{i=1}^{N} \sum_{n=1}^{N_{ij}} \bigl\|\tilde{\mathbf{e}}_{ijn} - \hat{\mathbf{e}}_{ijn}\bigr\|^{2}.
\]
Padded positions introduced to handle variable-length sequences within training batches are masked and excluded from the loss. Training is performed separately for each item~$j$, which adapts the latent representation to the behavioral vocabulary and the sequence characteristics of each task.

The output for respondent~$i$ and item~$j$ is the latent matrix
\[
\mathbf{C}_{ij} = (\mathbf{C}_{ij1}, \ldots, \mathbf{C}_{ijN_{ij}})^{\top} \in \mathbb{R}^{N_{ij} \times D},
\]
where each row summarizes a single action occurrence within its sequential context. The latent values are treated as fixed, observed covariates in the item response model, and $\bar{C}_{ijl}^{(d)}$ is computed from the rows of $\mathbf{C}_{ij}$ as defined in Section~\ref{sec:methodology}. The network architecture and hyperparameter settings are reported in Section~A of the Supplementary Material.


\subsection{The Representation-Learning Action-IRT Model with Spike-and-Slab Prior}\label{sec:entire_model}

The final stage links the latent action representations to response accuracy within an item response model. We extend the Rasch model with an additive action term that depends on the latent representations of the actions a respondent performed on the item. The extension preserves respondent ability and item difficulty parameters and quantifies the association between individual behavioral events and the probability of a correct response.

\subsubsection{Model Specification}\label{sec:model_spec}

For binary-scored items, $Y_{ij}$ is the recorded correct or incorrect outcome. For polytomously scored items (0--3), we set $Y_{ij}=1$ if the recorded score is 2 or 3, and $Y_{ij}=0$ otherwise. 
Let $\pi_{ij} = P(Y_{ij} = 1)$ denote the probability that respondent~$i$ answers item~$j$ correctly and let $I(\cdot)$ denote the indicator function. The representation-learning action-IRT model specifies the log-odds of a correct response as
\begin{equation}\label{eq:action_irt}
    \mathrm{logit}(\pi_{ij}) = \alpha_i + \beta_j + \sum_{l=1}^{L_j} \left( \sum_{d=1}^{D} \omega_{jl}^{(d)}\,\bar{C}_{ijl}^{(d)} \right) I(l \in A_{ij}),
\end{equation}
where $\alpha_i$ is the ability parameter for respondent~$i$, $\beta_j$ is the easiness (or -difficulty) parameter for item~$j$, and $\omega_{jl}^{(d)}$ is the action weight for unique action~$l$ in item~$j$ along latent dimension~$d$. The indicator $I(l \in A_{ij})$ restricts the outer sum to actions appearing in $A_{ij}$, and the contribution of action~$l$ to the log-odds equals $\sum_{d=1}^{D} \omega_{jl}^{(d)}\,\bar{C}_{ijl}^{(d)}$ when $l \in A_{ij}$ and zero otherwise.

Two features of this specification are important for interpretation. First, the covariate $\bar{\mathbf{C}}_{ijl}$ is respondent specific. Two respondents who perform the same action $l$ may have different representations because the LSTM encoder incorporates the sequential and temporal context in which the action occurs. Thus, the action index $l$ identifies what the respondent did, whereas $\bar{\mathbf{C}}_{ijl}$ captures aspects of how and when that action occurred within the response process. Second, $\boldsymbol{\omega}_{jl}$ is shared across respondents but specific to action $l$ within item $j$. It determines whether variation in the contextualized representation of that action is associated with response accuracy after accounting for respondent proficiency and item difficulty. Because the representations are learned without using response outcomes, these associations are estimated in the action-IRT stage rather than imposed during representation learning.

\subsubsection{Prior Specification and Variable Selection}\label{sec:prior}

The number of unique actions per item is large, and most actions are expected to have negligible associations with response accuracy conditional on ability and difficulty. We assign each action weight a spike-and-slab prior \citep{mitchellbeauchamp1988, george1993variable, ishwaran2005spike} that performs variable selection over the action space,
\begin{equation}\label{eq:spike_slab}
\omega_{jl}^{(d)} \mid \lambda_{jl}^{(d)} \sim
\begin{cases}
N(0,\,\tau^2), & \text{if } \lambda_{jl}^{(d)}=0
                 \quad (\text{spike}),\\[4pt]
N(0,\,\nu^2), & \text{if } \lambda_{jl}^{(d)}=1
                \quad (\text{slab}),
\end{cases}
\end{equation}
where $\tau^2 \ll \nu^2$ are the spike and slab variances. The spike component shrinks the weights of non-important actions toward zero. The slab component is diffuse and permits large weights for actions whose latent representations are associated with response accuracy. The latent binary indicator $\lambda_{jl}^{(d)}$ governs component membership and is assigned an independent Bernoulli prior,
\begin{equation}\label{eq:lambda_prior}
\lambda_{jl}^{(d)} \sim \mathrm{Bernoulli}(0.5),
\end{equation}
which places equal prior probability on importance and non-importance. The posterior distribution of $\lambda_{jl}^{(d)}$ drives the selection of important actions.

The priors on the remaining parameters are
\begin{equation}\label{eq:remaining_priors}
\alpha_i \mid \sigma_{\alpha}^{2} \sim N\!\left(0,\,\sigma_{\alpha}^{2}\right), \qquad \beta_j \sim N\!\left(0,\,\sigma_{\beta}^{2}\right), \qquad \sigma_{\alpha}^{2} \sim \mathrm{Inverse\text{-}Gamma}\!\left(a_0,\,b_0\right),
\end{equation}
where $\sigma_{\beta}^{2}$ is a fixed hyperparameter for the item difficulty prior. The values of $\sigma_{\beta}^{2}$, $a_0$, $b_0$, $\tau^2$, and $\nu^2$ are reported in Section~C of the Supplementary Material.

\subsubsection{Posterior Inference}\label{sec:post_inf}

Let $\boldsymbol{\theta} = \{\alpha_i, \beta_j, \omega_{jl}^{(d)}, \lambda_{jl}^{(d)}, \sigma_{\alpha}^{2}\}$ denote the full parameter collection, with indices ranging over $i = 1,\ldots,N$, $j = 1,\ldots,J$, $l = 1,\ldots,L_j$, and $d = 1,\ldots,D$. Let $\mathcal{O} = \{(i,j) : Y_{ij}\ \text{is observed}\}$ denote the set of observed respondent--item pairs. The likelihood, conditional on the fixed latent action representations $\{\bar{C}_{ijl}^{(d)}\}$, is
\begin{equation}\label{eq:likelihood}
L(\boldsymbol{\theta} \mid \mathbf{Y}) = \prod_{(i,j) \in \mathcal{O}} \pi_{ij}^{\,Y_{ij}} (1 - \pi_{ij})^{1 - Y_{ij}},
\end{equation}
where $\mathrm{logit}(\pi_{ij})$ is given by Equation~\eqref{eq:action_irt}. Combining the likelihood with the priors in Equations~\eqref{eq:spike_slab}, \eqref{eq:lambda_prior}, and~\eqref{eq:remaining_priors}, the joint posterior is
\begin{equation}\label{eq:joint_posterior}
p(\boldsymbol{\theta} \mid \mathbf{Y}) \;\propto\; L(\boldsymbol{\theta} \mid \mathbf{Y}) \prod_{i=1}^{N} p(\alpha_i \mid \sigma_{\alpha}^{2}) \prod_{j=1}^{J} p(\beta_j) \prod_{j,l,d} p(\omega_{jl}^{(d)} \mid \lambda_{jl}^{(d)})\, p(\lambda_{jl}^{(d)}) \cdot p(\sigma_{\alpha}^{2}).
\end{equation}
Direct sampling from Equation~\eqref{eq:joint_posterior} is intractable, and we employ a component-wise Markov chain Monte Carlo (MCMC) sampler in which each parameter block is updated from its full conditional distribution, following the Metropolis--Hastings-within-Gibbs strategy standard for item response models \citep{patz1999straightforward, fox2010bayesian}. The parameters $\alpha_i$, $\beta_j$, and $\omega_{jl}^{(d)}$ have no closed-form full conditionals under the logistic link and are updated by Metropolis--Hastings steps with Gaussian random-walk proposals. The inclusion indicators $\lambda_{jl}^{(d)}$ and the ability variance $\sigma_{\alpha}^{2}$ follow Bernoulli and inverse-gamma full conditional distributions and are updated by Gibbs sampling. The full conditional distributions, update scheme, proposal specifications, and convergence diagnostics are provided in Section~C of the Supplementary Material.

\subsubsection{Identification of Important Actions}\label{sec:ident_important}

The posterior samples of $\lambda_{jl}^{(d)}$ yield the posterior inclusion probability (PIP) for each action--dimension combination:
\[
\mathrm{PIP}_{jl}^{(d)} = P\!\left(\lambda_{jl}^{(d)} = 1 \mid \mathbf{Y}\right),
\]
estimated by the proportion of post-burn-in MCMC iterations in which $\lambda_{jl}^{(d)} = 1$. A high PIP indicates that the data favor the slab component for that action weight, providing evidence that the latent representation of the action is associated with response accuracy beyond the contributions of ability and difficulty.

We define the active dimension set for action~$l$ in item~$j$ as
\begin{equation}\label{eq:active_set}
    S_{jl} = \bigl\{d \in \{1, \ldots, D\} : \mathrm{PIP}_{jl}^{(d)} \geq 0.5\bigr\}.
\end{equation}
To translate a selected weight into an interpretable quantity, we summarize its contribution across the respondents who performed the action. Let $\mathcal{R}_{jl} = \{i : l \in A_{ij}\}$ denote that set of respondents. The average latent representation of action~$l$ along dimension~$d$ is
\begin{equation}\label{eq:action_mean}
    \bar{C}_{jl}^{(d)} = \frac{1}{|\mathcal{R}_{jl}|} \sum_{i \in \mathcal{R}_{jl}} \bar{C}_{ijl}^{(d)}.
\end{equation}

Let $t$ index the saved post-burn-in MCMC iterations. At iteration~$t$, the action effect is
\begin{equation}\label{eq:action_effect}
    \Delta_{jl}^{(t)} = \sum_{d \in S_{jl}} \omega_{jl}^{(d)(t)}\, I\!\left(\lambda_{jl}^{(d)(t)} = 1\right)\, \bar{C}_{jl}^{(d)},
\end{equation}
which retains the dimensions $d \in S_{jl}$ for which $\lambda_{jl}^{(d)(t)} = 1$. We write $\Delta_{jl}$ for the posterior distribution obtained by evaluating Equation~\eqref{eq:action_effect} at every saved iteration. We classify action~$l$ as an important action for item~$j$ when the 95\% highest posterior density (HPD) interval of $\Delta_{jl}$ excludes zero. The sign and magnitude of $\Delta_{jl}$ give the direction and strength of the contribution of the action to the probability of a correct response, and together with the posterior uncertainty they form the basis for the interpretation in Section~\ref{sec:real_data}. 

\section{Real Data Application Results}\label{sec:real_data}

The analysis includes respondents for whom the log process data and the scored response outcome are both available, which yields $N = 1{,}996$ respondents with the item-level sample sizes reported in Table~\ref{tab:items}. We set $D = 1$, which assigns one weight to each action--item combination and gives 2,025 action weights in total. Before model fitting, the latent action values were scaled within each item by centering at the item-level median and dividing by the interquartile range, so that action weights are comparable across items. Posterior samples were generated with the component-wise MCMC algorithm of Section~\ref{sec:post_inf}, run for 50,000 iterations with a burn-in of 10,000 and thinning by a factor of 10, which yields 4,000 saved samples. Metropolis--Hastings proposals were tuned so that median acceptance rates for $\alpha_i$ and $\beta_j$ fell between 0.2 and 0.4 \citep{roberts1997weak, roberts2001optimal}. Prior hyperparameters, proposal specifications, and sampler settings are reported in Section~C of the Supplementary Material.

Convergence was assessed with multiple chains initialized from dispersed starting values, together with trace plots and effective sample sizes. The Gelman--Rubin statistics \citep{gelman1992inference} were consistent with convergence, with no indication of disagreement between chains. Trace plots for $\alpha_i$ and $\beta_j$ showed stable mixing. Convergence of the 2,025 action weights was examined through summaries over all weights together with inspection of the weights of the selected actions, and both showed stable sampling. Trace plots, effective sample sizes, and Gelman--Rubin statistics for all parameter blocks are provided in Section~C of the Supplementary Material and in the accompanying GitHub repository.

\subsection{Summary of Selected Actions}\label{sec:summary}

Across the 14 PSTRE items, 126 actions were identified as important under the criterion that the 95\% HPD interval of $\Delta_{jl}$ excludes zero, which is 6.2\% of the 2,025 action--item combinations entering the analysis. Table~\ref{tab:selected_actions} reports, for each item, the number of unique action types in the model, the number identified as important, and the directional breakdown of the posterior action effect.

The number of selected actions per item ranges from 2 in PS2-3 (Book Order) to 16 in PS1-1 (Party Invitations Part~1) and PS1-6 (Tickets), with a median of 10. As a proportion of the action vocabulary of an item, selection ranges from 1.4\% in PS2-2 (Club Membership Part~2) to 14.9\% in PS1-4 (Sprained Ankle Part~1). Among the 126 selected actions, 72 have a positive posterior mean $\Delta_{jl}$ and 54 have a negative posterior mean. Selection is sparse in every item, and the composition of the selected set varies across items in ways examined in Section~\ref{sec:item_level}.

\begin{table}[!h]
\centering
\begin{tabular}{llrrrr}
\toprule
Item & Task & Total & Selected & Pos. & Neg. \\
\midrule
PS1-1 & Party Invitations (Part~1) & 119 & 16 & 10 &  6 \\
PS1-2 & Party Invitations (Part~2) & 141 & 10 &  5 &  5 \\
PS1-3 & CD Tally                   & 124 &  3 &  3 &  0 \\
PS1-4 & Sprained Ankle (Part~1)    &  47 &  7 &  3 &  4 \\
PS1-5 & Sprained Ankle (Part~2)    &  48 &  4 &  1 &  3 \\
PS1-6 & Tickets                    & 182 & 16 &  6 & 10 \\
PS1-7 & Class Attendance           & 200 & 11 &  5 &  6 \\
\midrule
PS2-1 & Club Membership (Part~1)   & 156 & 10 &  7 &  3 \\
PS2-2 & Club Membership (Part~2)   & 289 &  4 &  2 &  2 \\
PS2-3 & Book Order                 &  70 &  2 &  2 &  0 \\
PS2-4 & Meeting Room               & 206 & 14 &  7 &  7 \\
PS2-5 & Reply All                  & 112 & 10 &  5 &  5 \\
PS2-6 & Locate Email               & 146 & 11 &  9 &  2 \\
PS2-7 & Lamp Return                & 185 &  8 &  7 &  1 \\
\midrule
Total &                            & 2025 & 126 & 72 & 54 \\
\bottomrule
\end{tabular}
\caption{
Important actions identified for each PSTRE item. ``Total'' is the number of unique action types entering the model for the item. ``Selected'' is the number of actions whose 95\% highest posterior density interval of the action effect $\Delta_{jl}$ in Equation~\eqref{eq:action_effect} excludes zero. ``Pos.'' and ``Neg.'' count the selected actions with positive and with negative posterior mean $\Delta_{jl}$, corresponding to association with correct and with incorrect responses conditional on respondent ability and item difficulty.
}
\label{tab:selected_actions}
\end{table}

\subsection{Item-Level Analysis}\label{sec:item_level}

Two cross-item patterns emerge from the set of selected actions and motivate a unified framework for interpreting the action effects in the following case studies;
\begin{itemize}
\item {\bf Directional Balance} The ratio of positively to negatively selected actions varies with the functional structure of the task. Some items yield mainly positive effects, which indicates that correct problem solving in those tasks involves an identifiable set of productive behaviors. Examples include PS2-3 (Book Order, 2 positive and 0 negative) and PS2-6 (Locate Email, 9 positive and 2 negative). Other items yield a negative-dominant or balanced pattern, among them PS1-6 (Tickets, 6 positive and 10 negative) and PS1-5 (Sprained Ankle Part~2, 1 positive and 3 negative), which indicates that behaviors associated with incorrect responses also carry discriminating information.

\item {\bf Selection and Vocabulary Size} The number of selected actions does not increase monotonically with the size of the action vocabulary. PS2-2 (Club Membership Part~2) has the largest vocabulary at 289 unique actions and yields 4 selections, whereas PS1-1 and PS1-6, with vocabularies of 119 and 182, each yield 16. The density of behaviorally important actions therefore depends on the extent to which the interface of a task generates discriminating interactions rather than on vocabulary size.
\end{itemize}

The key quantity for interpretation is $\Delta_{jl}$, the posterior action effect defined in Equation~\eqref{eq:action_effect}. Two interpretive levels are distinguished throughout. At the first-order level, an action is selected as important when the 95\% HPD interval of $\Delta_{jl}$ excludes zero. The selection indicates that the latent representation of the action carries information about response accuracy beyond respondent ability $\alpha_i$ and item difficulty $\beta_j$. This inference is available for every selected action regardless of sequence length.

At the second-order level, the sign and magnitude of $\Delta_{jl}$ admit a process-level reading. The covariate $\bar{C}_{ijl}$ encodes the identity of action~$l$ together with the sequential context summarized by the LSTM hidden state and the associated timing (Section~\ref{sec:lstm_ae}), and $\Delta_{jl}$ measures the contribution of this composite representation to the log-odds of a correct response conditional on $\alpha_i$ and $\beta_j$. A positive $\Delta_{jl}$ indicates that the behavioral context in which the action typically occurs is associated with correct responses, and a negative value the reverse. The sign of $\Delta_{jl}$ need not follow raw action frequency. An action performed by correct and incorrect respondents in similar numbers may carry a negative effect when its typical sequential context accompanies less successful problem solving.

Second-order interpretation requires action sequences that are long and variable enough across respondents for the latent representations to encode context beyond the identity of individual actions \citep{chen2020continuous}. In items with extended multi-step sequences in which correct and incorrect respondents follow different trajectories, the LSTM encoder may capture contextual information such as exploration order and temporal pacing, and the sign of $\Delta_{jl}$ offers suggestive evidence about the nature of those differences. In items with short or structurally simple sequences, or in items where the trajectories of correct and incorrect respondents are similar, the latent representations may lack process-level signal, and any second-order reading should be treated as tentative.

The functional character of the selected actions divides the 14 PSTRE items into three types. In \textit{outcome-dominant} items, the leading discriminating action is a final-choice operation whose effect derives largely from its association with the scored outcome, and interpretation of such actions is restricted to the first-order level. In \textit{process-dominant} items, the selected actions are distributed across intermediate behavioral steps, which supports both levels of interpretation. In \textit{hybrid} items, outcome-proximal and process-level actions both appear among the selections, which permits first-order inference for all selected actions and second-order inference for the intermediate ones. Table~\ref{tab:item_types} assigns each item to a type.

\begin{table}[htb]
\centering
{\small
\begin{tabular}{cll}
\toprule
Label & Description & Items \\
\midrule
Outcome-dominant & Most selected actions are final-choice operations & PS1-3, PS1-4, PS1-5, PS1-7, PS2-3 \\
Process-dominant & Selected actions span intermediate process steps  & PS1-1, PS1-2, PS2-1, PS2-5, PS2-6 \\
Hybrid           & Outcome-proximal and process actions coexist     & PS1-6, PS2-2, PS2-4, PS2-7 \\
\bottomrule
\end{tabular}
}
\caption{
Typology of the 14 PSTRE items by the functional character of the actions selected as important. The type determines whether interpretation of the action effects is restricted to the first-order level or extends to the second-order level.
}
\label{tab:item_types}
\end{table}

The following subsections present one case study per type, PS1-3 (CD~Tally) for outcome-dominant items, PS1-1 (Party Invitations Part~1) for process-dominant items, and PS2-7 (Lamp Return) for hybrid items. Results for the remaining items are provided in Section~D of the Supplementary Material.

\subsubsection{Outcome-Dominant Items: CD Tally (PS1-3)}\label{sec:ps1_3}

In the CD~Tally task, respondents act as an employee updating the online inventory of a store. The interface comprises two linked environments, a web application and a spreadsheet. The spreadsheet presents each CD with its title, artist, genre, and release date in a sortable table. Respondents determine the number of Blues-genre CDs in the spreadsheet, switch to the web application, select that count from a numeric combobox, and submit the answer. Three actions were identified as important, the correct answer entry, a sorting step, and an interface switch.

\begin{table}[htp!]
\centering
\small
\begin{tabular}{clrc}
\toprule
Dir. & Action & $n$ & $\Delta$ (95\% HPD) \\
\midrule
$(+)$ & \texttt{combobox\_menulist\_index9}     & 492 & $\phantom{-}$6.2705 ($\phantom{-}$5.6520, $\phantom{-}$6.9219) \\
$(+)$ & \texttt{combobox\_sortablecol1\_index3} & 219 & $\phantom{-}$1.3804 ($\phantom{-}$0.1059, $\phantom{-}$2.7554) \\
$(+)$ & \texttt{toolbar\_webapp}               & 675 & $\phantom{-}$1.2928 ($\phantom{-}$0.2630, $\phantom{-}$2.2207) \\
\bottomrule
\end{tabular}
\caption{
Actions identified as important for PS1-3 (CD Tally). ``Dir.'' gives the sign of the posterior mean action effect and $n$ is the number of respondents who performed the action at least once. The column $\Delta$ reports the posterior mean of the action effect $\Delta_{jl}$ of Equation~\eqref{eq:action_effect} with the 95\% highest posterior density interval in parentheses, which measures the contribution of the average latent representation of the action to the log-odds of a correct response conditional on respondent ability and item difficulty. An action is listed when the interval excludes zero. Positive values correspond to association with correct responses and negative values to association with incorrect responses.
}
\label{tab:ps13_actions}
\end{table}

The action \path{combobox_menulist_index9} corresponds to selecting the correct Blues CD count. Its effect, $\Delta = 6.271$, is the largest of the three. The magnitude is inflated by quasi-complete separation, since respondents who perform this action are almost exclusively those who answer correctly, and it should not be compared with the other two effects. The direction remains interpretable.

The action \path{combobox_sortablecol1_index3} corresponds to sorting the spreadsheet by the genre column, which groups the Blues titles together and facilitates counting. Its effect, $\Delta = 1.380$, indicates that the procedural step carries information about response accuracy conditional on respondent ability and item difficulty. Deliberate sorting may reflect an organized solution strategy, and this process-level reading is treated as suggestive, in line with the first-order emphasis appropriate for outcome-dominant items.

The action \path{toolbar_webapp} represents switching from the spreadsheet to the web application. The transition is part of the task workflow, in which respondents determine the count in the spreadsheet and then navigate to the web application to select and submit the response. Its effect, $\Delta = 1.293$, indicates that the behavioral context surrounding the switch is associated with a correct response.

The three selected actions span the correct answer entry, a sorting step, and a task-required interface transition. The quasi-complete separation in \path{combobox_menulist_index9}, the sole final-choice action in this item, limits the extent to which the effects support process-level inference, as is characteristic of outcome-dominant items.

\subsubsection{Process-Dominant Items: Party Invitations (PS1-1)}\label{sec:ps1_1}

In the Party Invitations task, respondents act as a host who has sent party invitations by email and must read each incoming reply to determine whether the sender can attend. Each processed message is filed into the corresponding ``Can Come'' or ``Cannot Come'' subfolder, both of which reside within a parent ``Party'' folder, after which respondents submit the final answer. Sixteen actions were identified as important, the largest count among the 14 items. The action sequences are long and variable across respondents, with median length 16 and interquartile range 11 to 25, which supports both levels of interpretation.

\begin{table}[!h]
\centering
{\small
\begin{tabular}{clrc}
\toprule
Dir. & Action & $n$ & $\Delta$ (95\% HPD) \\
\midrule
$(+)$ & \texttt{button\_move\_validation}             &  109 & $\phantom{-}$2.2448 ($\phantom{-}$1.2599, $\phantom{-}$3.1033) \\
$(+)$ & \texttt{mail-drop\_u01a-cannotcomefolder}     &  695 & $\phantom{-}$1.4164 ($\phantom{-}$0.6800, $\phantom{-}$2.1399) \\
$(+)$ & \texttt{mail-drop\_u01a-cancomefolder}        &  913 & $\phantom{-}$1.3605 ($\phantom{-}$0.8843, $\phantom{-}$1.8507) \\
$(+)$ & \texttt{mail-drag\_u01a-item104}              &  830 & $\phantom{-}$0.9455 ($\phantom{-}$0.5227, $\phantom{-}$1.3479) \\
$(+)$ & \texttt{mail-viewed\_u01a-item102}            &  900 & $\phantom{-}$0.9377 ($\phantom{-}$0.3880, $\phantom{-}$1.4634) \\
$(+)$ & \texttt{mail-drag\_u01a-item101}              &  933 & $\phantom{-}$0.7767 ($\phantom{-}$0.2461, $\phantom{-}$1.2595) \\
$(+)$ & \texttt{mail-drag\_u01a-item102}              &  741 & $\phantom{-}$0.7738 ($\phantom{-}$0.0567, $\phantom{-}$1.5481) \\
$(+)$ & \texttt{mail-viewed\_u01a-item105}            &  885 & $\phantom{-}$0.6526 ($\phantom{-}$0.1754, $\phantom{-}$1.2187) \\
$(+)$ & \texttt{folder-viewed\_u01a-cannotcomefolder} &  534 & $\phantom{-}$0.6498 ($\phantom{-}$0.1142, $\phantom{-}$1.2189) \\
$(+)$ & \texttt{folder-viewed\_u01a-inboxfolder}      &  484 & $\phantom{-}$0.4249 ($\phantom{-}$0.2021, $\phantom{-}$0.6433) \\
$(-)$ & \texttt{mail-drop\_u01a-partyfolder}          &  128 & $-$0.4932 ($-$0.8365, $-$0.1660) \\
$(-)$ & \texttt{folder-viewed\_u01a-partyfolder}      &  326 & $-$0.5402 ($-$1.0255, $-$0.0018) \\
$(-)$ & \texttt{mail-viewed\_u01a-item202}            &  275 & $-$0.6587 ($-$1.4464, $-$0.0037) \\
$(-)$ & \texttt{toolbar\_mailapp}                     &   71 & $-$0.9085 ($-$1.6106, $-$0.2071) \\
$(-)$ & \texttt{mail-drag\_u01a-item202}              &  120 & $-$0.9937 ($-$1.7782, $-$0.1563) \\
$(-)$ & \texttt{mail-drag\_u01a-item201}              &  122 & $-$1.3575 ($-$2.2133, $-$0.5190) \\
\bottomrule
\end{tabular}
}
\caption{
Actions identified as important for PS1-1 (Party Invitations Part~1). Column definitions and the selection criterion are as in Table~\ref{tab:ps13_actions}. Message identifiers ending in 101, 102, 104, and 105 denote the replies to be classified, and those ending in 201 and 202 denote messages outside the classification targets.
}
\label{tab:ps11_actions}
\end{table}

At the first-order level, the 16 selected actions include task-relevant operations and behaviors outside the task requirements. The task-relevant operations are confirming placement (\path{button_move_validation}), dropping messages into the two destination folders, dragging the target messages, and reading target messages. The remaining behaviors are viewing or dragging extraneous messages, opening or dropping messages into the parent Party folder, and switching interfaces (\path{toolbar_mailapp}). The selection of both categories indicates that the model captures information beyond the identity of the correct classification operations.

At the second-order level, positive effects concentrate on the core classification steps. The largest are confirming message placement ($\Delta = 2.245$), dropping messages into the Cannot Come folder ($\Delta = 1.416$), and dropping messages into the Can Come folder ($\Delta = 1.361$), which indicates that the sorting operations are associated with a correct response. Dragging the three primary target messages (\path{mail-drag_u01a-item104}, $\Delta = 0.946$; \path{mail-drag_u01a-item101}, $\Delta = 0.777$; \path{mail-drag_u01a-item102}, $\Delta = 0.774$) and reading target messages (\path{mail-viewed_u01a-item102}, $\Delta = 0.938$; \path{mail-viewed_u01a-item105}, $\Delta = 0.653$) also carry positive effects, which indicates that engagement with the relevant messages accompanies correct task completion. The actions \path{folder-viewed_u01a-cannotcomefolder} ($\Delta = 0.650$) and \path{folder-viewed_u01a-inboxfolder} ($\Delta = 0.425$) carry smaller positive effects, consistent with an association between organized folder navigation and correct responses.

Negative effects concentrate on engagement directed away from the classification task. Dragging extraneous messages (\path{mail-drag_u01a-item201}, $\Delta = -1.358$; \path{mail-drag_u01a-item202}, $\Delta = -0.994$) and viewing an extraneous message (\path{mail-viewed_u01a-item202}, $\Delta = -0.659$) carry the largest negative effects, which indicates that engagement with messages outside the classification targets is associated with incorrect responses. Switching to the mail application (\path{toolbar_mailapp}, $\Delta = -0.909$) carries a negative effect of similar size. Dropping messages into the parent Party folder (\path{mail-drop_u01a-partyfolder}, $\Delta = -0.493$) and opening that folder (\path{folder-viewed_u01a-partyfolder}, $\Delta = -0.540$) are associated with misdirected classification attempts.

The selected actions thus contrast engagement with the target messages and use of the destination folders, associated with correct responses, against engagement with extraneous messages and misdirected navigation, associated with incorrect responses.

\subsubsection{Hybrid Items: Lamp Return (PS2-7)}\label{sec:ps2_7}

In the Lamp Return task, respondents have ordered a desk lamp online, received it in the wrong color, and must use the company website to exchange it for the color originally ordered. The procedure requires navigating to the customer-service page and completing a return request form. The form requires a return authorization number, which respondents must first request through a simulated email client, retrieve from the subsequent reply, enter into the form, and submit with the completed request. Eight actions were identified as important, spanning intermediate procedural steps and a final confirmatory operation. The action sequences are of moderate length, with median 15 and interquartile range 6 to 24, and the trajectories of correct and incorrect respondents differ enough to support both levels of interpretation.

\begin{table}[!h]
\centering
\small
\begin{tabular}{clrc}
\toprule
Dir. & Action & $n$ & $\Delta$ (95\% HPD) \\
\midrule
$(+)$ & \texttt{textlink\_u023-pg2\_txt21\_self}   & 265 & $\phantom{-}$2.9700 ($\phantom{-}$1.6828, $\phantom{-}$4.3001) \\
$(+)$ & \texttt{button\_u023-pg2\_txt22}           & 523 & $\phantom{-}$2.6306 ($\phantom{-}$2.0668, $\phantom{-}$3.2000) \\
$(+)$ & \texttt{mail-viewed\_u023-item305}          & 612 & $\phantom{-}$1.3857 ($\phantom{-}$0.4728, $\phantom{-}$2.2566) \\
$(+)$ & \texttt{toolbar\_back-btn}                & 735 & $\phantom{-}$1.2777 ($\phantom{-}$0.6018, $\phantom{-}$1.9751) \\
$(+)$ & \texttt{textbox-onfocus\_u023-pg2\_txt19}  & 528 & $\phantom{-}$1.1382 ($\phantom{-}$0.7491, $\phantom{-}$1.5808) \\
$(+)$ & \texttt{keypress2}                        & 354 & $\phantom{-}$0.8201 ($\phantom{-}$0.3446, $\phantom{-}$1.2832) \\
$(+)$ & \texttt{combobox\_u023-pg2\_menu1\_index1} & 559 & $\phantom{-}$0.2106 ($\phantom{-}$0.0059, $\phantom{-}$0.4047) \\
$(-)$ & \texttt{textlink\_u023-default\_txt5}      & 474 & $-$0.6309 ($-$1.1537, $-$0.1058) \\
\bottomrule
\end{tabular}
\caption{
Actions identified as important for PS2-7 (Lamp Return). Column definitions and the selection criterion are as in Table~\ref{tab:ps13_actions}. 
}
\label{tab:ps27_actions}
\end{table}

The seven positively selected actions cover each stage of the return procedure, and the estimated effects follow its procedural order. Requesting the authorization email (\path{textlink_u023-pg2_txt21_self}, $\Delta = 2.970$), the first necessary step, and submitting the completed request (\path{button_u023-pg2_txt22}, $\Delta = 2.631$), the concluding step, carry the largest positive effects. Reviewing the authorization reply (\path{mail-viewed_u023-item305}, $\Delta = 1.386$) and navigating back to the return form (\path{toolbar_back-btn}, $\Delta = 1.278$) follow, which indicates that the cross-application navigation required to retrieve and transfer the authorization number is associated with successful task execution. Activating the authorization-number input field (\path{textbox-onfocus_u023-pg2_txt19}, $\Delta = 1.138$) and entering the number by keypress (\path{keypress2}, $\Delta = 0.820$) are also associated with correct responses. Selecting the return reason (\path{combobox_u023-pg2_menu1_index1}, $\Delta = 0.211$) carries the smallest effect, consistent with its position as an early procedural step completed before the more discriminating later stages. 

The action \path{textlink_u023-default_txt5} ($\Delta = -0.631$) is the sole negative selection. Its interface function could not be identified from the available log metadata, and we offer no behavioral interpretation. The distribution of the selected actions across the procedural sequence indicates a behavioral pathway toward correct task completion and exemplifies the hybrid type, in which discriminating information spans intermediate steps and a final confirmatory operation.

\section{Model Validation Using Simulation Studies}\label{sec:simul}

We conducted a simulation study to evaluate the variable-selection performance of the representation-learning action-IRT model. The framework treats the LSTM output as a fixed covariate, and the simulation targets the action-IRT stage. The observed respondent--item structure, action-occurrence patterns, and latent action values are held at their empirical values, and only the binary responses are regenerated. The design inherits the covariate structure, missingness pattern, and sparsity level of the empirical analysis.

The data-generating model was calibrated to the empirical fit of Section~\ref{sec:real_data}, retaining 14 PSTRE items, 1,996 respondents, 2,025 action--item combinations, and $D = 1$, with the covariates $\bar{C}_{ijl}$ held at their empirical values. The number of unique actions per item and the counts of true important actions are those in Table~\ref{tab:selected_actions}.

True parameter values were taken from the empirical posterior. Respondent ability and item difficulty were fixed at their posterior means, $\alpha_i^{\mathrm{true}} = \mathbb{E}(\alpha_i \mid \mathbf{Y})$ and $\beta_j^{\mathrm{true}} = \mathbb{E}(\beta_j \mid \mathbf{Y})$. The 126 action--item combinations identified as important in the empirical analysis were designated true important actions, each assigned $\lambda_{jl}^{\mathrm{true}} = 1$ with true weight equal to the slab-conditional posterior mean of $\omega_{jl}$. The remaining combinations were assigned $\lambda_{jl}^{\mathrm{true}} = 0$ with weights drawn independently from the spike component.

For each replication, binary responses were generated for the observed respondent--item pairs as $Y_{ij} \sim \mathrm{Bernoulli} \bigl(\mathrm{logistic}(\eta_{ij})\bigr)$, with
\[
\eta_{ij} = \alpha_i^{\mathrm{true}}+\beta_j^{\mathrm{true}} + \sum_{l\in A_{ij}}\omega_{jl}^{\mathrm{true}}\,\bar{C}_{ijl},
\]
where $A_{ij}$ is the set of unique actions performed by respondent~$i$ on item~$j$. Five replications were carried out with distinct random seeds under the MCMC sampler and posterior-sampling settings of the empirical analysis, reported in Section~\ref{sec:real_data} and in Section~C of the Supplementary Material. 

Variable-selection performance was evaluated using the posterior inclusion probability $\mathrm{PIP}_{jl} = P(\lambda_{jl} = 1 \mid \mathbf{Y})$, estimated as the proportion of saved post-burn-in iterations in which $\lambda_{jl} = 1$. Because $D = 1$, the action-level PIP coincides with the single-dimension PIP. We report the area under the receiver operating characteristic curve (AUC), computed with the true inclusion indicator as the binary label and PIP as the continuous score.

\begin{figure}[htb!]
\centering
\includegraphics[width=0.65\linewidth]{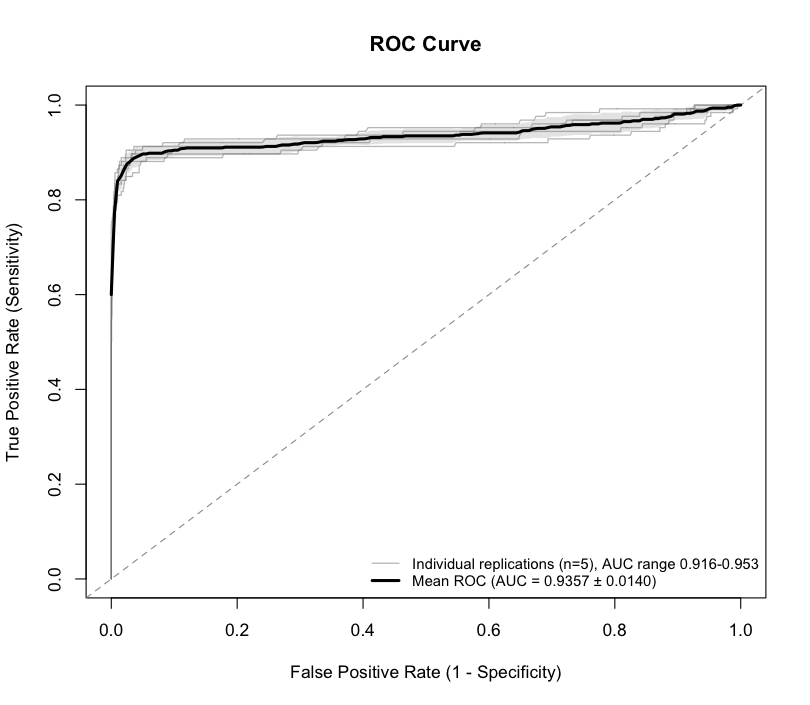}
\caption{
Receiver operating characteristic curves for selection based on posterior inclusion probabilities, shown for each of the five simulation replications together with their mean. The true inclusion indicator $\lambda_{jl}^{\mathrm{true}}$ serves as the binary label and $\mathrm{PIP}_{jl}$ as the score, and each curve traces the true positive rate against the false positive rate as the selection threshold varies over the 2,025 action--item combinations. The diagonal corresponds to selection at chance level. AUC values are reported in the legend.
}
\label{fig:roc_curve}
\end{figure}

Figure~\ref{fig:roc_curve} shows the ROC curves for the five replications. The action-level AUC averaged 0.9357 (SD 0.0140; range 0.9160--0.9530), indicating that the posterior inclusion probability reliably discriminates truly important actions from non-important ones across replications. The model recovers the empirically designated important actions when the observed action structure and LSTM latent values are preserved.

\section{Conclusion}\label{sec:conc}

This paper proposed a framework that couples representation learning with model-based selection for identifying behaviorally important actions in problem-solving process data. Action representations are learned from the raw log sequences through two components, a hybrid Word2Vec embedding that captures the sequential co-occurrence of actions together with the compositional structure of action labels, and an LSTM autoencoder that encodes each action within its sequential context. Neither component requires manual feature construction. 
The per-action latent values enter an extended Rasch model with a spike-and-slab prior so that estimation and variable selection proceed simultaneously, and every action--item combination receives a posterior inclusion probability with selection conditional on respondent ability and item difficulty. 
The representation-learning and selection components operate on action sequences rather than on features specific to PSTRE tasks, and the framework applies in principle to other sequential process data in which identifying behaviorally important events is of interest.

Applied to log process data from 14 PSTRE items in the United States sample of the first PIAAC cycle, the framework selected 126 important actions among 2,025 action--item combinations, a selection rate of 6.2\%. Item-level case studies supported a typological distinction among outcome-dominant, process-dominant, and hybrid items. In outcome-dominant items such as CD~Tally (PS1-3), behavioral discrimination concentrates at the final-choice step. In process-dominant items such as Party~Invitations (PS1-1), discriminating actions are distributed across intermediate procedural steps, and the posterior action effects distinguish behaviors associated with correct responses from those associated with incorrect ones. In hybrid items such as Lamp~Return (PS2-7), the selected actions span the procedural sequence. The action effect $\Delta_{jl}$ reflects the sequential context of each action rather than its identity alone, and these patterns capture behavioral distinctions that response outcomes and raw action frequencies would not reveal.

Several limitations qualify these findings. Second-order interpretation of the action effects is limited for outcome-dominant items in which behavioral variation concentrates at the final-choice step and the sequences of correct and incorrect respondents are not differentiated. Actions performed by only one response group are subject to quasi-complete separation, and their effect magnitudes should not be compared with those of other actions. The empirical analysis relies on a single latent dimension ($D = 1$), which may compress information that a higher-dimensional representation would preserve. The embedding model and the autoencoder are fitted separately for each item, which limits cross-item information sharing. The plug-in treatment of the latent action values does not propagate uncertainty from the dimension-reduction stage into the IRT posterior, and the simulation study holds these values fixed, leaving the representation-learning stages outside its scope.

The framework offers a reproducible procedure for extracting behavioral evidence from process data within an established measurement tradition, and the identified actions may serve as a data-driven complement to expert-designed process indicators for post hoc strategy assessment. Future work could examine sensitivity to the choice of latent dimension and prior specification, compare alternative action representations, extend the model to joint cross-item estimation, and incorporate other behavioral signals toward a fuller account of problem-solving competence in digital assessment environments.

\section*{Acknowledgments}

This work was partially supported by the National Research Foundation of Korea [grant number NRF-2021S1A3A2A03088949, RS-2023-00217705, and RS-2024-00333701; Basic Science Research Program awarded to I.H.J.], the Basic Science Research Program through the National Research Foundation of Korea (NRF) funded by the Ministry of Education [grant number RS-2025-25426200, awarded to J. Park], the ICAN (ICT Challenge and Advanced Network of HRD) support program [grant number RS-2023-00259934, awarded to I.H.J.], supervised by the IITP (Institute of Information \& Communications Technology Planning \& Evaluation), and the Ministry of Trade, Industry, and Energy (MOTIE), Korea, under the project ``Industrial Technology Infrastructure Program'' [RS-2024-00466693, awarded to I.H.J.]. Correspondence should be addressed to Ick Hoon Jin, Department of Applied Statistics, Department of Statistics and Data Science, Yonsei University, Seoul, Republic of Korea. E-Mail: ijin@yonsei.ac.kr. 

\section*{Data Availability}
The datasets generated during and/or analyzed during the current study are available in the GESIS, \url{https://doi.org/10.4232/1.12955}. The preprocessing scripts, sampler code, and simulation files accompanying this paper are maintained at \url{https://github.com/P-JuNYeonG/action-irt}.

\section*{AI Use Statement}
Large language models were used at three points in the preparation of this manuscript. Claude Sonnet 4.5 was used for the description standardization step of the preprocessing pipeline, as described in Sections~\ref{sec:preprocessing_pipeline} and~\ref{sec:llm_preprocessing}. Figure~\ref{fig:framework} was drafted with ChatGPT (GPT-5.5, medium reasoning setting). Claude Opus 5.0 was used to check for typographical and grammatical errors. The authors reviewed all output and take full responsibility for the content of the manuscript.

\newpage
\bibliographystyle{fixed-Chicago}
\bibliography{reference}

@book{oecd2013piaac,
  author    = {{OECD}},
  title     = {{OECD} Skills Outlook 2013: First Results from the Survey of Adult Skills},
  year      = {2013},
  publisher = {OECD Publishing},
  address   = {Paris},
  doi       = {10.1787/9789264204256-en},
  url       = {https://doi.org/10.1787/9789264204256-en}
}

@article{greiff2016understanding,
  title  = {Understanding students' performance in a computer-based assessment of complex problem solving: {An} analysis of behavioral data from computer-generated log files},
  author  = {Greiff, Samuel and Niepel, Christoph and Scherer, Ronny and Martin, Romain},
  journal = {Computers in Human Behavior},
  volume  = {61},
  pages   = {36--46},
  year    = {2016},
  publisher = {Elsevier}
}

@article{kroehne2018conceptualize,
  title   = {How to conceptualize, represent, and analyze log data from technology-based assessments? {A} generic framework and an application to questionnaire items},
  author  = {Kroehne, Ulf and Goldhammer, Frank},
  journal = {Behaviormetrika},
  volume  = {45},
  number  = {2},
  pages   = {527--563},
  year    = {2018},
  publisher = {Springer}
}

@article{goldhammer2021byproduct,
  title   = {From byproduct to design factor: {On} validating the interpretation of process indicators based on log data},
  author  = {Goldhammer, Frank and Hahnel, Carolin and Kroehne, Ulf and Zehner, Fabian},
  journal = {Large-Scale Assessments in Education},
  volume  = {9},
  number  = {1},
  pages   = {20},
  year    = {2021},
  publisher = {Springer}
}

@techreport{maddox2023uses,
  author      = {Maddox, Bryan},
  title       = {The Uses of Process Data in Large-Scale Educational Assessments},
  institution = {OECD},
  address     = {Paris},
  year        = {2023},
  month       = jan,
  doi         = {10.1787/5d9009ff-en},
  note        = {{OECD} Education Working Papers No. 286}
}

@book{oecd2012literacy,
  title   = {{Literacy, Numeracy and Problem Solving in Technology-Rich Environments: Framework for the OECD Survey of Adult Skills}},
  author  = {{OECD}},
  year    = {2012},
  publisher = {OECD Publishing}
}

@techreport{piaac2009pstre,
  author      = {{PIAAC Expert Group in Problem Solving in Technology-Rich Environments}},
  title       = {{PIAAC} Problem Solving in Technology-Rich Environments: {A} Conceptual Framework},
  institution = {OECD},
  year        = {2009},
  type        = {OECD Education Working Papers},
  number      = {36},
  address     = {Paris},
  publisher   = {OECD Publishing},
  doi         = {10.1787/220262483674},
  url         = {https://doi.org/10.1787/220262483674}
}

@book{organisation2019beyond,
  title   = {{Beyond Proficiency: Using Log Files to Understand Respondent Behaviour in the Survey of Adult Skills}},
  author  = {{OECD}},
  year    = {2019},
  publisher = {OECD}
}

@incollection{goldhammer2020analysing,
  title     = {{Analysing Log File Data From PIAAC}},
  author    = {Goldhammer, Frank and Hahnel, Carolin and Kroehne, Ulf},
  booktitle = {{Large-Scale Cognitive Assessment: Analyzing PIAAC Data}},
  pages     = {239--269},
  year      = {2020},
  publisher = {Springer International Publishing Cham}
}

@article{stadler2020first,
  title   = {First among equals: {Log} data indicates ability differences despite equal scores},
  author  = {Stadler, Matthias and Hofer, Sarah and Greiff, Samuel},
  journal = {Computers in Human Behavior},
  volume  = {111},
  pages   = {106442},
  year    = {2020},
  publisher = {Elsevier}
}

@article{eichmann2020exploring,
  title   = {Exploring behavioural patterns during complex problem-solving},
  author  = {Eichmann, Beate and Greiff, Samuel and Naumann, Johannes and Brandhuber, Liene and Goldhammer, Frank},
  journal = {Journal of Computer Assisted Learning},
  volume  = {36},
  number  = {6},
  pages   = {933--956},
  year    = {2020},
  publisher = {Wiley Online Library}
}

@article{he2021leveraging,
  title   = {Leveraging process data to assess adults' problem-solving skills: {Using} sequence mining to identify behavioral patterns across digital tasks},
  author  = {He, Qiwei and Borgonovi, Francesca and Paccagnella, Marco},
  journal = {Computers \& Education},
  volume  = {166},
  pages   = {104170},
  year    = {2021},
  publisher = {Elsevier}
}

@inproceedings{he2015identifying,
  title        = {Identifying feature sequences from process data in problem-solving items with n-grams},
  author       = {He, Qiwei and von Davier, Matthias},
  booktitle    = {Quantitative Psychology Research: The 79th Annual Meeting of the Psychometric Society, Madison, Wisconsin, 2014},
  pages        = {173--190},
  year         = {2015},
  organization = {Springer}
}

@incollection{he2016analyzing,
  title     = {{Analyzing Process Data From Problem-solving Items With N-grams: Insights From a Computer-based Large-scale Assessment}},
  author    = {He, Qiwei and von Davier, Matthias},
  booktitle = {{Handbook of Research on Technology Tools for Real-World Skill Development}},
  pages     = {750--777},
  year      = {2016},
  publisher = {IGI Global Scientific Publishing}
}

@article{tang2020latent,
  title   = {Latent feature extraction for process data via multidimensional scaling},
  author  = {Tang, Xueying and Wang, Zhi and He, Qiwei and Liu, Jingchen and Ying, Zhiliang},
  journal = {Psychometrika},
  volume  = {85},
  number  = {2},
  pages   = {378--397},
  year    = {2020},
  publisher = {Cambridge University Press \& Assessment}
}

@article{tang2021exploratory,
  title   = {An exploratory analysis of the latent structure of process data via action sequence autoencoders},
  author  = {Tang, Xueying and Wang, Zhi and Liu, Jingchen and Ying, Zhiliang},
  journal = {British Journal of Mathematical and Statistical Psychology},
  volume  = {74},
  number  = {1},
  pages   = {1--33},
  year    = {2021},
  publisher = {Wiley Online Library}
}

@techreport{he2019using,
  author      = {He, Q. and Borgonovi, F. and Paccagnella, M.},
  title       = {Using process data to understand adults' problem-solving behaviour in the {Programme for the International Assessment of Adult Competencies} ({PIAAC}): {Identifying} generalised patterns across multiple tasks with sequence mining},
  institution = {OECD Publishing},
  series      = {OECD Education Working Papers},
  number      = {205},
  year        = {2019},
  address     = {Paris},
  doi         = {10.1787/650918f2-en},
  url         = {https://doi.org/10.1787/650918f2-en}
}

@article{wang2023subtask,
  title   = {Subtask analysis of process data through a predictive model},
  author  = {Wang, Zhi and Tang, Xueying and Liu, Jingchen and Ying, Zhiliang},
  journal = {British Journal of Mathematical and Statistical Psychology},
  volume  = {76},
  number  = {1},
  pages   = {211--235},
  year    = {2023},
  publisher = {Wiley Online Library}
}

@article{zhou2024investigating,
  title   = {Investigating response behavior through {TF-IDF} and {Word2Vec} text analysis: {A} case study of {PISA} 2012 problem-solving process data},
  author  = {Zhou, Jing and Ye, Zhanliang and Zhang, Sheng and Geng, Zhao and Han, Ning and Yang, Tao},
  journal = {Heliyon},
  volume  = {10},
  number  = {16},
  pages   = {e35945},
  year    = {2024},
  publisher = {Elsevier}
}

@book{rasch1960probabilistic,
  title     = {Probabilistic Models for Some Intelligence and Attainment Tests},
  author    = {Rasch, Georg},
  year      = {1960},
  publisher = {Danish Institute for Educational Research},
  address   = {Copenhagen}
}

@inproceedings{mikolov2013distributed,
  title     = {Distributed Representations of Words and Phrases and Their Compositionality},
  author    = {Mikolov, Tomas and Sutskever, Ilya and Chen, Kai and Corrado, Greg S. and Dean, Jeffrey},
  booktitle = {Advances in Neural Information Processing Systems 26 ({NIPS} 2013)},
  year      = {2013},
  pages     = {3111--3119}
}

@inproceedings{mikolov2013efficient,
  title     = {Efficient Estimation of Word Representations in Vector Space},
  author    = {Mikolov, Tomas and Chen, Kai and Corrado, Greg and Dean, Jeffrey},
  booktitle = {Proceedings of the International Conference on Learning Representations ({ICLR}), Workshop Track},
  year      = {2013}
}

@article{bojanowski2017enriching,
  author  = {Bojanowski, Piotr and Grave, Edouard and Joulin, Armand and Mikolov, Tomas},
  title   = {Enriching Word Vectors with Subword Information},
  journal = {Transactions of the Association for Computational Linguistics},
  year    = {2017},
  volume  = {5},
  pages   = {135--146},
  doi     = {10.1162/tacl_a_00051}
}

@article{hochreiter1997long,
  title   = {Long Short-Term Memory},
  author  = {Hochreiter, Sepp and Schmidhuber, J{\"u}rgen},
  journal = {Neural Computation},
  volume  = {9},
  number  = {8},
  pages   = {1735--1780},
  year    = {1997},
  doi     = {10.1162/neco.1997.9.8.1735}
}

@article{hinton2006reducing,
  title   = {Reducing the Dimensionality of Data with Neural Networks},
  author  = {Hinton, Geoffrey E. and Salakhutdinov, Ruslan R.},
  journal = {Science},
  volume  = {313},
  number  = {5786},
  pages   = {504--507},
  year    = {2006},
  doi     = {10.1126/science.1127647}
}

@article{mitchellbeauchamp1988,
  title   = {{Bayesian} Approach to Variable Selection in linear Regression},
  author  = {Mitchell, Tom J. and Beauchamp, John J.},
  journal = {Journal of the american statistical association},
  volume  = {83},
  number  = {404},
  pages   = {1023--1032},
  year    = {1988},
  doi     = {10.1080/01621459.1988.10478694}
}

@article{george1993variable,
  title   = {Variable Selection via {Gibbs} Sampling},
  author  = {George, Edward I. and McCulloch, Robert E.},
  journal = {Journal of the American Statistical Association},
  volume  = {88},
  number  = {423},
  pages   = {881--889},
  year    = {1993},
  doi     = {10.1080/01621459.1993.10476353}
}

@inproceedings{barkan2016item2vec,
  title        = {{Item2Vec}: Neural item embedding for collaborative filtering},
  author       = {Barkan, Oren and Koenigstein, Noam},
  booktitle    = {2016 {IEEE} 26th International Workshop on Machine Learning for Signal Processing ({MLSP})},
  pages        = {1--6},
  year         = {2016},
  organization = {IEEE}
}

@inproceedings{grbovic2015ecommerce,
  author    = {Grbovic, Mihajlo and Radosavljevic, Vladan and Djuric, Nemanja and Bhamidipati, Narayan and Savla, Jaikit and Bhagwan, Varun and Sharp, Doug},
  title     = {E-commerce in Your Inbox: {Product} Recommendations at Scale},
  booktitle = {Proceedings of the 21th {ACM SIGKDD} International Conference on Knowledge Discovery and Data Mining},
  pages     = {1809--1818},
  year      = {2015},
  doi       = {10.1145/2783258.2788627},
  url       = {https://doi.org/10.1145/2783258.2788627}
}

@article{pardos2020university,
  title     = {A university map of course knowledge},
  author    = {Pardos, Zachary A and Nam, Andrew Joo Hun},
  journal   = {{PLOS ONE}},
  volume    = {15},
  number    = {9},
  pages     = {e0233207},
  year      = {2020},
  publisher = {Public Library of Science San Francisco, CA USA}
}

@inproceedings{srivastava2015unsupervised,
  title        = {Unsupervised learning of video representations using {LSTM}s},
  author       = {Srivastava, Nitish and Mansimov, Elman and Salakhudinov, Ruslan},
  booktitle    = {International Conference on Machine Learning},
  pages        = {843--852},
  year         = {2015},
  organization = {PMLR}
}

@misc{malhotra2016lstm,
  title         = {{LSTM}-based Encoder-Decoder for Multi-sensor Anomaly Detection},
  author        = {Malhotra, Pankaj and Ramakrishnan, Anusha and Anand, Gaurangi and Vig, Lovekesh and Agarwal, Puneet and Shroff, Gautam},
  year          = {2016},
  eprint        = {1607.00148},
  archivePrefix = {arXiv},
  primaryClass  = {cs.LG},
  url           = {https://arxiv.org/abs/1607.00148}
}

@article{patz1999straightforward,
  title   = {A straightforward approach to {Markov} chain {Monte Carlo} methods for item response models},
  author  = {Patz, Richard J and Junker, Brian W},
  journal = {Journal of Educational and Behavioral Statistics},
  volume  = {24},
  number  = {2},
  pages   = {146--178},
  year    = {1999},
  publisher = {Sage Publications}
}

@article{van2007hierarchical,
  title   = {A hierarchical framework for modeling speed and accuracy on test items},
  author  = {Van der Linden, Wim J},
  journal = {Psychometrika},
  volume  = {72},
  number  = {3},
  pages   = {287--308},
  year    = {2007},
  publisher = {Springer}
}

@article{wang2015mixture,
  title   = {A mixture hierarchical model for response times and response accuracy},
  author  = {Wang, Chun and Xu, Gongjun},
  journal = {British Journal of Mathematical and Statistical Psychology},
  volume  = {68},
  number  = {3},
  pages   = {456--477},
  year    = {2015},
  publisher = {Wiley Online Library}
}

@article{ishwaran2005spike,
  title   = {Spike and slab variable selection: {Frequentist} and {B}ayesian strategies},
  author  = {Ishwaran, Hemant and Rao, J. Sunil},
  journal = {The Annals of Statistics},
  year    = {2005},
  volume  = {33},
  number  = {2},
  pages   = {730--773}
}

@article{chen2022advantages,
  title   = {Advantages of spike and slab priors for detecting differential item functioning relative to other {Bayesian} regularizing priors and frequentist lasso},
  author  = {Chen, Siyuan Marco and Bauer, Daniel J and Belzak, William M and Brandt, Holger},
  journal = {Structural Equation Modeling: A Multidisciplinary Journal},
  volume  = {29},
  number  = {1},
  pages   = {122--139},
  year    = {2022},
  publisher = {Taylor \& Francis}
}

@misc{hwangbo2026tutorial,
  title        = {Analyzing Process Data from Computer-Based Assessments: {A} Tutorial on Preprocessing, Feature Extraction, and Model-Based Inference},
  author       = {Hwangbo, Daeun and Park, Junyeong and Jeon, Minjeong and
                  Jin, Ick Hoon},
  year         = {2026},
  eprint       = {2604.16900},
  archivePrefix = {arXiv},
  primaryClass = {stat.AP},
  url          = {https://arxiv.org/abs/2604.16900}
}

@techreport{anthropic2025sonnet,
  title       = {{Claude Sonnet 4.5 System Card}},
  author      = {{Anthropic}},
  institution = {Anthropic},
  year        = {2025},
  month       = {September},
  note        = {Revised December 3, 2025},
  url         = {https://www.anthropic.com/claude-sonnet-4-5-system-card}
}

@article{goldhammer2014time,
  author  = {Goldhammer, F. and Naumann, J. and Stelter, A. and T{\'o}th, K. and R{\"o}lke, H. and Klieme, E.},
  title   = {The time on task effect in reading and problem solving is moderated by task difficulty and skill: {Insights} from a computer-based large-scale assessment},
  journal = {Journal of Educational Psychology},
  volume  = {106},
  number  = {3},
  pages   = {608--626},
  year    = {2014},
  doi     = {10.1037/a0034716},
  url     = {https://doi.org/10.1037/a0034716}
}

@article{lundgren2020within,
  title     = {Within-item response processes as indicators of test-taking effort
               and motivation},
  author    = {Lundgren, Erik and Ekl{\"o}f, Hanna},
  journal   = {Educational Research and Evaluation},
  volume    = {26},
  number    = {5-6},
  pages     = {275--301},
  year      = {2020},
  publisher = {Taylor \& Francis}
}

@article{costa2023exploring,
  title     = {Exploring the relationship between process data and contextual variables among {Scandinavian} students on {PISA} 2012 mathematics tasks},
  author    = {Costa, Denise Reis and Chen, Chia-Wen},
  journal   = {Large-Scale Assessments in Education},
  volume    = {11},
  number    = {1},
  pages     = {5},
  year      = {2023},
  publisher = {Springer}
}

@article{park2025analysis,
  title     = {Analysis of log data from an international online educational assessment system: {A} multi-state survival modeling approach to reaction time between and across action sequence},
  author    = {Park, Jina and Jin, Ick Hoon and Jeon, Minjeong},
  journal   = {Psychometrika},
  volume    = {90},
  number    = {4},
  year      = {2025},
  doi       = {10.1017/psy.2025.10043},
  publisher = {Cambridge University Press}
}

@book{fox2010bayesian,
  author    = {Fox, Jean-Paul},
  title     = {{Bayesian} Item Response Modeling: Theory and Applications},
  series    = {Statistics for Social and Behavioral Sciences},
  publisher = {Springer},
  address   = {New York, NY},
  year      = {2010},
  edition   = {1},
  doi       = {10.1007/978-1-4419-0742-4},
  isbn      = {978-1-4419-0742-4}
}

@book{deboeck2004explanatory,
  editor    = {De Boeck, Paul and Wilson, Mark},
  title     = {Explanatory Item Response Models: {A} Generalized Linear and Nonlinear Approach},
  series    = {Statistics for Social and Behavioral Sciences},
  publisher = {Springer},
  address   = {New York, NY},
  year      = {2004},
  doi       = {10.1007/978-1-4757-3990-9},
  isbn      = {978-1-4757-3990-9}
}

@article{tang2025path,
  title={A path signature perspective of process data feature extraction},
  author={Tang, Xueying and Liu, Jingchen and Ying, Zhiliang},
  journal={British Journal of Mathematical and Statistical Psychology},
  volume={78},
  number={3},
  pages={939--964},
  year={2025},
  publisher={Wiley Online Library}
}

@article{ulitzsch2022using,
  title={Using sequence mining techniques for understanding incorrect behavioral patterns on interactive tasks},
  author={Ulitzsch, Esther and He, Qiwei and Pohl, Steffi},
  journal={Journal of Educational and Behavioral Statistics},
  volume={47},
  number={1},
  pages={3--35},
  year={2022},
  publisher={Sage Publications Sage CA: Los Angeles, CA}
}

@article{chen2020continuous,
  title={A continuous-time dynamic choice measurement model for problem-solving process data},
  author={Chen, Yunxiao},
  journal={Psychometrika},
  volume={85},
  DOI={10.1007/s11336-020-09734-1},
  number={4},
  pages={1052--1075},
  year={2020}
}

@article{ulitzsch2021combining,
  title={Combining clickstream analyses and graph-modeled data clustering for identifying common response processes},
  author={Ulitzsch, Esther and He, Qiwei and Ulitzsch, Vincent and Molter, Hendrik and Nichterlein, Andr{\'e} and Niedermeier, Rolf and Pohl, Steffi},
  journal={Psychometrika},
  volume={86},
  number={1},
  pages={190--214},
  year={2021}
}

@article{zhang2023identifying,
  title={Identifying problem-solving solution patterns using network analysis of operation sequences and response times},
  author={Zhang, Maoxin and Andersson, Bj{\"o}rn},
  journal={Educational Assessment},
  volume={28},
  number={3},
  pages={172--189},
  year={2023},
  publisher={Taylor \& Francis}
}

@article{stadler2019taking,
  title={Taking a closer look: {An} exploratory analysis of successful and unsuccessful strategy use in complex problems},
  author={Stadler, Matthias and Fischer, Frank and Greiff, Samuel},
  journal={Frontiers in Psychology},
  volume={10},
  pages={777},
  year={2019},
  publisher={Frontiers Media SA}
}

@article{xu2020latent,
  title={A latent topic model with {Markov} transition for process data},
  author={Xu, Haochen and Fang, Guanhua and Ying, Zhiliang},
  journal={British Journal of Mathematical and Statistical Psychology},
  volume={73},
  number={3},
  pages={474--505},
  year={2020},
  publisher={Wiley Online Library}
}

@article{han2022sequential,
  title={A sequential response model for analyzing process data on technology-based problem-solving tasks},
  author={Han, Yuting and Liu, Hongyun and Ji, Feng},
  journal={Multivariate Behavioral Research},
  volume={57},
  number={6},
  pages={960--977},
  year={2022},
  publisher={Taylor \& Francis}
}

@article{fang2020latent,
  title={Latent theme dictionary model for finding co-occurrent patterns in process data},
  author={Fang, Guanhua and Ying, Zhiliang},
  journal={Psychometrika},
  volume={85},
  number={3},
  pages={775--811},
  year={2020}
}

@article{roberts1997weak,
  title={Weak convergence and optimal scaling of random walk {Metropolis} algorithms},
  author={Roberts, Gareth O and Gelman, Andrew and Gilks, Walter R},
  journal={The Annals of Applied Probability},
  volume={7},
  number={1},
  pages={110--120},
  year={1997},
  publisher={Institute of Mathematical Statistics}
}

@article{gelman1992inference,
  title={Inference from iterative simulation using multiple sequences},
  author={Gelman, Andrew and Rubin, Donald B},
  journal={Statistical Science},
  volume={7},
  number={4},
  pages={457--472},
  year={1992},
  publisher={Institute of Mathematical Statistics}
}

@article{roberts2001optimal,
  title={Optimal scaling for various {Metropolis-Hastings} algorithms},
  author={Roberts, Gareth O and Rosenthal, Jeffrey S},
  journal={Statistical Science},
  volume={16},
  number={4},
  pages={351--367},
  year={2001},
  doi     = {10.1214/ss/1015346320}
}

@article{zhang2023accurate,
  title   = {Accurate assessment via process data},
  author  = {Zhang, Susu and Wang, Zhi and Qi, Jitong and Liu, Jingchen and Ying, Zhiliang},
  journal = {Psychometrika},
  volume  = {88},
  number  = {1},
  pages   = {76--97},
  year    = {2023},
  doi     = {10.1007/s11336-022-09880-8}
}

@article{yun2025discovering,
  title   = {Discovering action insights from large-scale assessment log data using machine learning},
  author  = {Yun, Minyoung and Jeon, Minjeong and Yang, Heyoung},
  journal = {Scientific Reports},
  volume  = {15},
  pages   = {30412},
  year    = {2025},
  doi     = {10.1038/s41598-025-14802-6}
}

@article{zoanetti2010interactive,
  author  = {Zoanetti, N.},
  title   = {Interactive computer based assessment tasks: {How} problem-solving process data can inform instruction},
  journal = {Australasian Journal of Educational Technology},
  volume  = {26},
  number  = {5},
  year    = {2010},
  doi     = {10.14742/ajet.1053},
  url     = {https://doi.org/10.14742/ajet.1053}
}

@inproceedings{scarlatos2022process,
  title     = {Process-{BERT}: A framework for representation learning on educational process data},
  author    = {Scarlatos, Alexander and Brinton, Christopher and Lan, Andrew},
  booktitle = {Proceedings of the 15th International Conference on Educational Data Mining},
  year      = {2022},
  address   = {Durham, UK},
  publisher = {International Educational Data Mining Society}
}

\end{document}